\documentstyle[preprint,aps,eqsecnum,epsfig,amstex]{revtex}

\makeatletter           % Switch to floating figures
\@floatstrue
\def\figure{\let\@capwidth\columnwidth\@float{figure}}
\let\endfigure\end@float
\@namedef{figure*}{\let\@capwidth\textwidth\@dblfloat{figure}}
\@namedef{endfigure*}{\end@dblfloat}
\def\table{\let\@capwidth\columnwidth\@float{table}}
\let\endtable\end@float
\@namedef{table*}{\let\@capwidth\textwidth\@dblfloat{table}}
\@namedef{endtable*}{\end@dblfloat}

\def\la{\label}
\makeatother

\newcommand{\nc}{\newcommand}

\nc\pcite[1]{\protect{\cite{#1}}}
\nc{\bea}{\begin{eqnarray}}
\nc{\eea}{\end{eqnarray}}
\nc{\beqa}{\begin{eqnarray}}
\nc{\eeqa}{\end{eqnarray}}
\nc{\NCS}{N_{\rm CS}}
\nc{\lsim}{\mbox{\raisebox{-.6ex}{~$\stackrel{<}{\sim}$~}}}
\nc{\gsim}{\mbox{\raisebox{-.6ex}{~$\stackrel{>}{\sim}$~}}}
\nc{\mD}{m_{\rm D}}
\nc{\meq}{m_{\rm eq}}
\nc{\mth}{m_{\rm th}}
\nc{\Teq}{T_{\rm eq}}
\nc{\avphi}{{\phi^2_{\rm av}}}
\nc{\avphio}{\phi^2_{\rm av,0}{}}
\nc{\avphiC}{\phi^2_{\rm av,C}{}}
\nc{\avphiI}[1]{\phi^2_{{\rm av},#1}}
\nc{\avchange}{\langle | \Delta \avphi / \Delta t | \rangle}
\nc{\dispavchg}{\left\langle \left| \frac{\Delta \avphi}{\Delta
t}\right|\right\rangle}
\nc{\bitspace}{\hspace{0.3in}}
\nc{\sigmaE}{\sigma_{\rm el}}

\nc{\fig}{Fig.~}

\def\d{{\bf d}}

\begin {document}

\tightenlines

%%%%%%%%%%%%%%%%%%%%%%%%%%%%%%%%%%%%%%%%%%%%%%%%%%%%%%%%%%%%%%%%%%%%%%%%%%%%%%%

\preprint {UW/PT 00-01, NORDITA 2000/79HE}

\title{ Electroweak Bubble Nucleation, Nonperturbatively } 

\author {Guy D. Moore}

\address
    {%
Department of Physics,
University of Washington, 
Seattle WA 98195-1560 USA	
    }%

\author {Kari Rummukainen}

\address
    {%
NORDITA,
Blegdamsvej 17, 
DK-2100 Copenhagen \O,
Denmark
    }%

\date {June 2000} %Let's be realistic here

\maketitle
\vskip -20pt

\begin {abstract}%
{%
We present a lattice 
method to compute bubble nucleation rates at radiatively
induced first order phase transitions, in high temperature, weakly
coupled field theories, nonperturbatively.  A generalization of Langer's
approach, it makes no recourse to
saddle point expansions and includes completely the dynamical prefactor.
We test the technique by applying it to the electroweak phase transition
in  the minimal standard model, at an unphysically small Higgs mass
which gives a reasonably strong phase transition ($\lambda/g^2 =0.036$,
which corresponds to $m_H/m_W = 0.54$ at tree level but does not
correspond to a positive physical Higgs mass when radiative effects of
the top quark are included), and compare the results to older
perturbative and other estimates.  While two loop perturbation theory
slightly under-estimates the strength of the transition measured by the
latent heat, it over-estimates the amount of supercooling by a factor of
2. 
}%
\end {abstract}

\thispagestyle{empty}

%%%%%%%%%%%%%%%%%%%%%%%%%%%%%%%%%%%%%%%%%%%%%%%%%%%%%%%%%%%%%%%%%%%%%%%%%%%%%%%

\section {Introduction}

Electroweak baryogenesis provides one of the best motivated and most
testable mechanisms for the origin of the cosmological baryon number
abundance.  However, several ingredients are missing before we can make 
quantitative predictions.  One set of needed ingredients 
is particle physics inputs.  For instance, it is difficult to say much
about electroweak matter at temperatures $T \sim 100$GeV, where
electroweak symmetry is ``restored''\footnote{There is no qualitative
distinction between the ``symmetric'' and ``broken'' electroweak phases, 
which are in fact analytically connected,
and the symmetry is never truly ``broken'' even in vacuum; but when the
phase transition is reasonably strong their quantitative behavior is
very different.  We will use the ``symmetric'' and ``broken''
terminology because it is convenient and widespread.}
and baryogenesis may occur, when we
still do not know the Higgs mass.  It is also absolutely necessary to
know what other light ($m\lsim 150$GeV) scalars there are, what
couplings they have with the Higgs boson(s), and what CP violation is
operative at electroweak energy and temperature scales.  Answering these 
questions will require new experimental results, and we will not have
more to say about that here.

However, even if we knew the complete electroweak theory we could not at 
this time make accurate predictions of what baryon number would be
cosmologically produced, because we do not have a complete set of
reliable computational tools for studying the electroweak phase
transition and the physical processes responsible for baryogenesis.
This is best illustrated by briefly reviewing what the scenario is, and
which aspects we do or do not currently have good control of.

Assuming a standard thermal history for the early universe back
to\footnote{See \pcite{Joyce} for an interesting
discussion of what could happen if this assumption were not true.}
 $T \sim 100$GeV, 
electroweak baryogenesis appears to be possible only if there is a
fairly strong first order electroweak phase transition
\cite{old_Shap,KLRS_results}.  The electroweak phase transition, if there 
is one, is radiatively induced, and determining its order and strength
is a difficult problem with a long history.
In summary, perturbation theory proves to be of some limited use
when the phase transition is strong, but a reliable calculation of the
strength of the electroweak phase transition requires a nonperturbative
lattice calculation.  The equipment for performing an accurate lattice
calculation now exists, using either a 3-dimensional effective theory
\cite{KLRS_results,FKRS,KLRS,Oapaper,KLRS_cross,endpt} or 
4-dimensional SU(2) gauge + Higgs theory \cite{desy,Budapest}, so this
part of the problem is solved.

If the electroweak phase transition is first order, then the universe
will remain in the ``symmetric'' phase even after it is no longer
thermodynamically favored.  How deeply it supercools is the topic of
this paper and we will return to it momentarily.  After sufficiently
deep supercooling, critical bubbles of the broken phase form at a
cosmologically relevant rate, expand, and coalesce, completing the phase 
transition.\footnote{It has been argued that the phase transition can
also proceed by coalescence of ``subcritical bubbles'' with almost no
supercooling \pcite{Gleiser}.  We feel our technique and results,
presented here, put this idea to rest for the phase transition strength we
are interested in.}  It is the expansion of these bubbles into the
symmetric phase which is expected to generate the baryon number.
Specifically, the moving phase interface (bubble wall) 
can inject a CP violating flux
of particles into the symmetric phase \cite{CKN_or_something}, 
where baryon number violation is
efficient \cite{Kuzmin,ArnoldMcLerran}.  Recently, the efficiency of the 
baryon number violation has been pinned down fairly accurately
\cite{Bodeker,Bodek_paper,BMR,AY2_long,Bodek_higgs}.  However, neither the
expansion of bubbles into the symmetric phase, nor the generation and
propagation of CP violating particle fluxes, can yet be calculated with
much precision or confidence, though there has been recent progress on
both problems \cite{bubV,Schmidt1,Schmidt2,ClineKainulain2000}.

The bubble nucleation rate enters the final baryon number asymmetry by
determining the amount of supercooling which occurs.  The more
supercooling, the greater the free energy difference between the phases, 
and the faster phase interfaces propagate.  This in turn would mean a
larger injected CP violating flux (since the flux must vanish in
equilibrium in a CPT respecting theory).  However, it would also mean
that the phase interface would more quickly catch back up with particles 
it injected into the symmetric phase, which could have the reverse
effect.  The detailed dependence on supercooling and bubble wall
velocity may be complicated, see for instance
\cite{Heckler,ClineKainulain2000}.  How deep the supercooling proceeds
can also be important because the universe heats during the phase
transition, as the latent heat of the
symmetric phase is released.  We find below that the supercooling is
less than in perturbation theory, while the latent heat is sometimes
more; so this effect is more important than one might have anticipated.
For the parameters we will study, there is enough latent heat, and
little enough supercooling, that the universe reheats to the equilibrium 
temperature $\Teq$ before all space has converted to the broken phase.
(The remaining space would convert much more slowly, as the expansion of 
the universe continues to absorb heat from the plasma.)

We are interested in a regime where the bubble nucleation rate is
extremely small.  This is because the phase transition completes when
the bubble nucleation rate is, very roughly, around one bubble per
Hubble volume per Hubble time.  But at $T \sim 100$GeV, a Hubble time is 
$t_{\rm hubble} \sim H^{-1} \sim m_{\rm pl}/T^2 \sim 10^{17}/T$; so the
rate of bubble nucleations must be $\sim (10^{-17} T)^4 = 
10^{-68}T^4 = e^{-157} T^4$.  
A more careful calculation, accounting for how much time the phase
transition takes to complete, shows that the nucleation rate must be
about $e^{-106} T^4$.  When the
nucleation rate is so small, it is ``almost'' a thermodynamical
quantity, set by the free energy of the ``critical bubble.''  This gives us
a hint at how to determine it on the lattice; we must
determine the free energy of a critical bubble.  However, deciding what
precisely this means and how to go about doing it, and relating the
result to the real time rate for a rare process, require some work.  In
this paper we present a quantitative approach to address this problem,
and we carry out our program for the minimal standard model,
at zero Weinberg angle and an unphysical Higgs mass.  Clearly this case
will not be of direct physical interest.  However, it allows us to
determine how well the technique works, and to compare the bubble
nucleation rate to what we would get using one of several less
rigorous methods, such as a perturbative calculation of the critical
bubble free energy or a ``thin wall'' treatment from either perturbative or
nonperturbative inputs.  We find that the ``thin wall'' treatment gives
a reasonable answer but is not extremely accurate, while the
perturbative approach is quite bad unless Higgs field wave function
corrections are included.

Dynamical processes in a first order phase transition have been
studied before with lattice simulations in the 3-dimensional Ising model
\cite{Stauffer}.  However, in these studies the parameters of the
simulations were chosen so that the nucleation rate was relatively
large, so that the nucleation timescale was, at most, only few orders
of magnitude larger than the microscopic interaction timescale.  Thus,
the nucleation process could be observed simply by waiting for the
nucleation from a metastable to a stable state to happen during a
straightforward standard (real-time) simulation.  This is also the
case for the recent work of Borsanyi et.~al.~\cite{fate_class_vac}.

However, in this work we are interested in extremely strongly
suppressed nucleation.  Indeed, very slow nucleation rates are a 
quite common characteristic of the first order phase transitions in
Nature: fundamentally, this is due to the fact that the external
parameters which drive the transition (temperature, say) usually vary
on several orders of magnitude longer timescales than the microscopic
interaction scale.  This physical situation is out of reach of any
numerical real-time simulation method relying on spontaneous
appearance of bubbles in the metastable phase.  On the other hand, the
method presented in this work can be applied to an almost arbitrarily
slow nucleation.  This method has also been used with the 3-dimensional
cubic anisotropy model \cite{cubic}.

An outline of the paper is as follows.  In Section \ref{Plan}, we
present the approach and discuss the obstacles.  The discussion does
not rely in any way on the specifics of the electroweak nucleation
problem, except that it can be considered as a problem in classical
statistical mechanics (both in the thermodynamics and in the dynamics
of the system).  Subsections \ref{why_rare} and \ref{real_time},
together with subsection
\ref{multi_subsec}, are the most important parts of the paper to
understand; we encourage the reader to concentrate on them. 
In Section \ref{statmech}, we review why it is true that the physics of
the electroweak phase transition can be considered, both
thermodynamically and dynamically, as a classical statistical mechanics
problem.  This section is a review of previous literature, included
mostly to make the paper self-contained.  In
Section \ref{Numerics} we present the numerical tools we use and the
details of the calculation.  It ends with a presentation of our
results.  Section \ref{otherways} presents a number of alternative,
``more perturbative'' and less numerically intensive ways to try to
determine the bubble nucleation rate, most of which
have previously been used in the literature.  We systematically compare
these approaches to the nonperturbatively determined result, to analyze
the reliability of the other approaches.  Of these, the most
trustworthy are the thin wall approximation using nonperturbative inputs
(surface tension and latent heat), and two loop perturbation theory
including Higgs field wave function corrections.  Each of these approaches
makes errors of order $20\%$; other approaches, including those most
widely used in the literature, give results off by almost a factor of
2.  Finally, Section \ref{Conclusion} presents our conclusions. 

%------------------------------------------------------------------
\section{Strategy to Determine Nucleation Rate}
\la{Plan}

The next section will show why the calculation of the bubble nucleation
rate at the electroweak phase transition is a problem in classical
statistical mechanics.  In this section we will assume this to be the
case, and discuss the strategy for 
solving this statistical mechanics problem.  
The basic idea is that
nucleation from the metastable symmetric phase to the stable broken
phase is limited by the rarity of ``critical bubble'' configurations
which lie in between.  A ``critical bubble'' is, roughly, a
configuration which in the medium term\footnote{We will distinguish
three time scales.  The short time scale is the longest time scale of
typical thermalization processes in either phase, $\sim 1/g^4 T$.  The
long time scale is the time scale for nucleation to occur, $\sim
e^{100}/T$.  By the medium term we mean on a time scale well
separated from the short and long time scales.  At many points in our
discussion there will be ambiguities of order the ratio of two of these
time scales; but for strongly exponentially suppressed nucleation
problems such ambiguities are tiny.}  
is about equally likely to evolve towards the broken and
homogeneous symmetric phases.  An outline of the strategy is:
\begin{enumerate}
\item choose a measurable which will distinguish which field
	configurations are near the metastable symmetric phase, which
	are near the broken phase, and which are near the critical
	bubble; 
\item evaluate the probability to be in the (exponentially rare)
	critical bubble configurations; 
\item determine how quickly a ``critical bubble'' evolves towards one of 
	the (meta)stable phases;
\item determine the ``dynamical prefactor,'' which tells what fraction
	of imputed ``critical bubbles'' really represent midpoints on a
	trajectory carrying the metastable symmetric phase to the stable
	broken one. 
\end{enumerate}
This section elucidates what we mean by each of the above, 
and why the whole approach is possible.  Our strategy is similar to
Langer's classic method \cite{Langer}, except that the saddle point
treatment of the ``critical'' configurations is replaced by a
Monte Carlo calculation, and we take care to treat completely the
microscopic dynamics during ``barrier crossing.''
Essentially the same technique
has already been applied to determine the broken phase ``sphaleron
rate'' in \cite{broken1,broken_nonpert}, but we will not assume the
reader is familiar with those papers.

Everything we say in this section is generic to first order phase
transitions of liquid-gas type, ie.~where there is no breaking of a
global symmetry but the phases can be distinguished by the value of
the volume average of a single, scalar local measurable.  (However,
this method is also fully applicable to transitions exhibiting a real
global symmetry breaking.)  In our case the volume averaged observable
will be the average of the $\mbox{(length)}^2$ of the Higgs field,
$\avphi \equiv (1/V) \int d^3 x 2 \Phi^\dagger \Phi$.  If it makes the
reader more comfortable she can think of water and gas below the
critical temperature, with the density as a measurable distinguishing the
phases and pressure as a control parameter, or of a ferromagnet below 
the Curie temperature, with the magnetization as a measurable and an
applied external field as a control parameter, rather than the
electroweak model with $\avphi$ as a measurable and temperature as a
control parameter.  The big difference from the liquid gas system is that 
we know how to simulate the microscopic physics accurately even away
from the second order endpoint, so
Monte Carlo techniques applied to a first principles microscopic
description of the thermodynamics can give reliable results.

With suitable generalizations the method described here can be applied
to almost any metastable state decay problem, provided that 
\begin{enumerate}
\item both the
thermodynamics and the 
real time evolution of the system are amenable to numerical analysis,
\item there is an observable which can unambiguously distinguish the
phases and can resolve the potential barrier between the them, and
\item the potential barrier is large and the tunnelling rate is small.
\end{enumerate}
The first two conditions are quite generic, and if the third condition
is not satisfied (for example, in ``quenching'' type problems), then
one can make straightforward real-time simulation of the decay without
relying on the special methods described here.

\subsection{General picture of homogeneous nucleation after weak supercooling}
\la{why_rare}

\begin{figure}
\centerline{\epsfxsize=5in\epsfbox{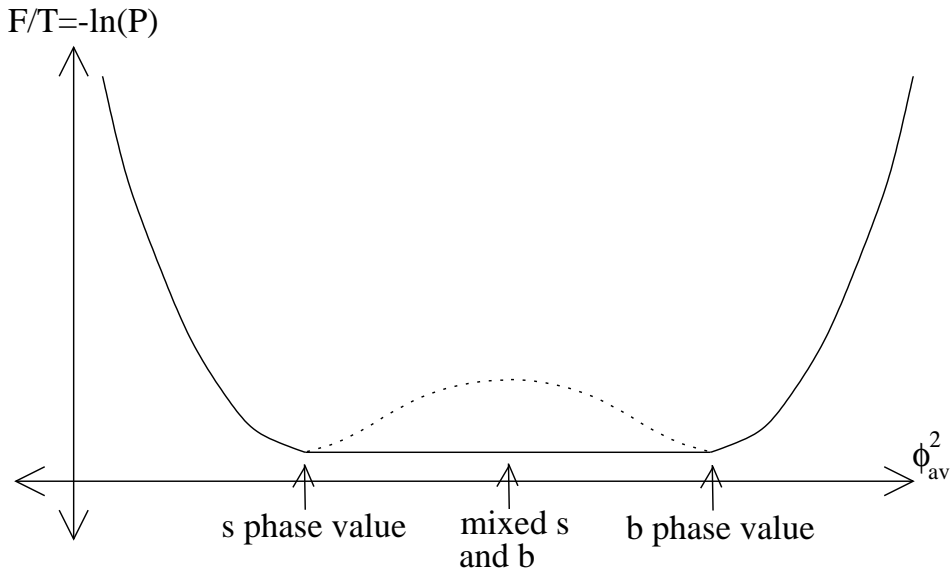}}
\vspace{0.2in}
\caption{\la{maxwell} Cartoon of how the constrained
free energy = $-\log$ (probability of $\avphi$) varies with
$\avphi$ at the equilibrium temperature in a large volume.  The
vertical axis gives minus the log of the fraction of states in the
canonical ensemble with the given value of $\avphi$.  The dotted line
gives the free energy of a {\em spatially homogeneous} configuration
with that value of $\avphi$; the truly most probable configurations at
intermediate values are mixed phase configurations.}
\end{figure}

Fix the ratio of the scalar self-coupling and gauge couplings, 
$\lambda / g^2$, to a value where there is a first order electroweak
phase transition (or fix $T$ to a value below the Curie temperature for
a ferromagnet, or $T$ below the critical temperature for a liquid-gas
system), and ask how the canonical ensemble is distributed
close to the critical temperature.  In particular, consider how the
constrained free energy $F(\avphi)$
%\begin{equation}
%  F(\avphi)/T = -\ln \int {\cal D}(A_i,\Phi)\, e^{-H/T}\, \delta[\avphi - \avphi(\Phi)] 
%\end{equation}
varies as a function of 
\begin{equation}
\avphi \equiv \frac{1}{V} \int d^3 x 
	( 2 \Phi^\dagger \Phi(x) - {\rm counterterm} ) \, ,
\end{equation}
where the counterterm is needed to subtract UV divergences so the
whole is well defined.  
%(For the full definition of the action we use, see Section \ref{statmech}.)
We could consider any other measurable
which has UV finite variance, but $\avphi$ will prove particularly
convenient below.  In a very large volume, at $\Teq$ (or zero external
magnetic field in a ferromagnet, or the boiling pressure in a
liquid-gas system), the constrained free energy density dependence on
$\avphi$ will qualitatively resemble that of Fig.~\ref{maxwell}.  The
constrained free energy means $-T$ times the log of the weight of
configurations in the canonical ensemble which have the specific value
of $\avphi$.  Very low or very high values of $\avphi$ are extremely
rare; but values intermediate between the two bulk phases have free
energy  per volume equal to those of the bulk phases.  This is because,
besides the ``expensive way'' of getting an intermediate value of
$\avphi$ --- having $\Phi^\dagger \Phi$ equal the desired value
homogeneously through the volume of interest --- there is a ``cheaper
way,'' which is to have part of the volume be in one phase and the
rest in the other phase.  Then the disfavored intermediate values of
$\Phi^\dagger \Phi$ are only achieved in an interface between the
regions, whose volume does not scale extensively with the system
volume.  (Note that the fact that intermediate values of $\Phi^\dagger
\Phi$ are disfavored, is exactly the statement that there is a first
order phase transition with different values of $\Phi^\dagger\Phi$ in
the two phases.)\footnote{%
Strictly speaking, when we talk about
spatial variations of $\Phi^\dagger\Phi$ or spatially homogeneous
$\Phi^\dagger\Phi$ we actually mean a ``coarse-grained'' quantity:
normally $\Phi^\dagger\Phi$ fluctuates wildly from point to point even
in the pure symmetric or broken phases (it is, indeed, UV divergent).
The coarse-grained $\Phi^\dagger\Phi$ is obtained by averaging over
length scales $\approx$ bulk correlation length $\xi$.  This is
equivalent to integrating out spatial momenta larger than $\xi^{-1}$.
In the pure bulk phases the coarse-grained $\Phi^\dagger\Phi$ is
almost homogeneous by construction, and non-homogeneous only in the
mixed phase.%
}

What if we ask about a smaller volume, where the amount of space in the
interface between extensive phases is not negligible?  A qualitative
cartoon of the answer is given in Fig.~\ref{maxwell2}.  There is a free
energy ``barrier'' between the two phases due to the free energy cost of 
the interface separating the two phases; but it is much lower than it
would be if we had to stick with spatially homogeneous intermediate
states.  The barrier is 
roughly the surface tension of the interface times its
area, which scales as the length squared of the box, while an extensive
quantity would scale as length cubed.

\begin{figure}
\centerline{\epsfxsize=6in\epsfbox{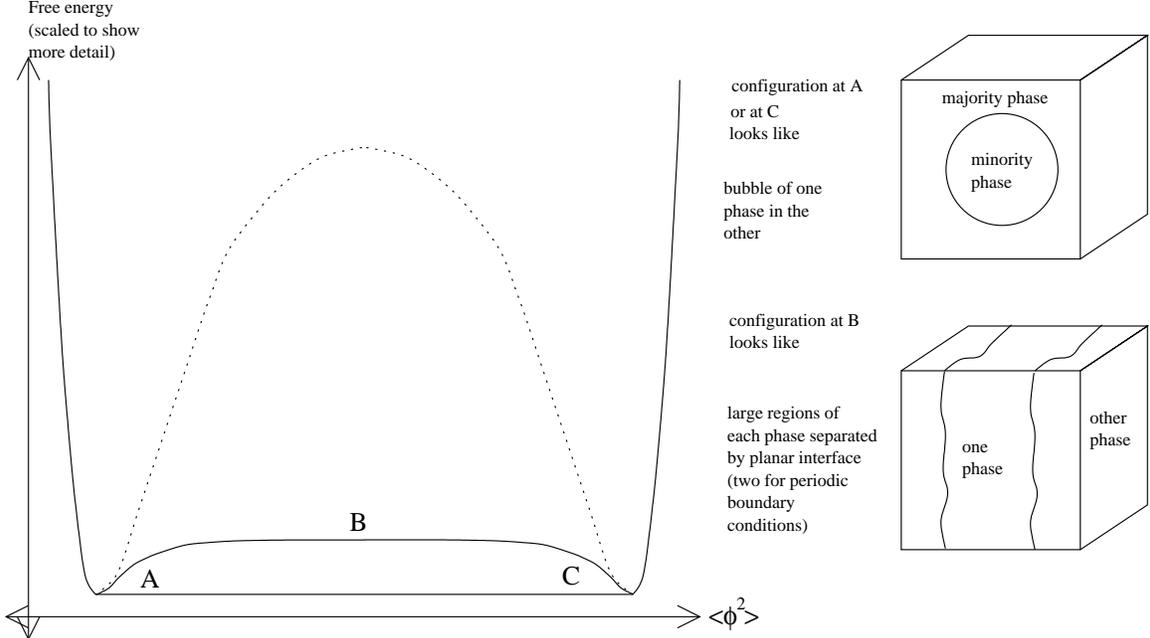}}
\caption{\la{maxwell2} Same as Fig.~\protect{\ref{maxwell}}, but in a
finite volume where the free energy of the interface between phases is
not considered negligible.  The free energy of a mixed phase state is
higher than either pure phase because of the surface tension of the
phase boundary.  The figure also illustrates the physical appearances of 
the states which dominate the ensemble at intermediate values of
$\avphi$.}
\end{figure}

The free energy at values of $\avphi$ between the stable phases gives us 
information about the free energy cost of mixed phase configurations; in 
particular, the free energy near the symmetric phase tells about the
cost to have a small bubble of the broken phase in the symmetric
phase.  To see the value of this, in determining the rate of bubble
nucleation, we now discuss how the picture changes when we change the
temperature.  The short answer is that one should ``tip''
Fig.~\ref{maxwell2}, adding a linear in $\avphi$ term to the free
energy.  In fact, for a special measurable this statement is exact, as
we now discuss in some detail.

In the 3-D effective theory approximation we are working
in (see next section), 
a variation $\delta T$ of the temperature corresponds to a change
$\delta m_{HT}^2$ in the thermal Higgs mass squared.  The size of the
change can be read off from Eq.~(\ref{m_of_T}).  Henceforth we will only 
talk about changing $m_{HT}^2$. 
The way that one determines the constrained free
energy plots we have been discussing is that one finds the probability
that a configuration drawn from the canonical ensemble has the value of
$\avphi$ of interest.  If we are interested in the free energy as a
function of an operator ${\cal O}$, we want to know
\begin{equation}
\frac{F( {\cal O}_0 ) }{T} = 
	- \ln \int {\cal D}(A_i,\Phi) \exp( -H/T ) \; \;
	\delta \Big( {\cal O}(A,\Phi) - {\cal O}_0 \Big) \, ,
\end{equation}
up to an overall constant which is uninteresting (and depends on how we
define the normalization of the path integral).  For the special
case that our operator is $\avphi$, the constrained free energy has an
extremely convenient property.  Observe from Eq. (\ref{3D-theory}) 
that the way that $m_{HT}^2$ enters the Hamiltonian is
\begin{eqnarray}
H & = & H_{m=m_0} + \frac{m_{HT}^2 - m_0^2}{2} \int d^3 x 
	2 \Phi^\dagger \Phi(x) \, , \\
& = & H_{m=m_0} + \frac{m_{HT}^2-m_0^2}{2} V \avphi \, , \nonumber
\end{eqnarray}
with $m_0^2$ any particular value we might choose.
The constrained free energy as a function of $\avphi$ is
\begin{equation}
\frac{F( \avphio )}{T} = - \ln \int {\cal D} (A_i,\Phi)
	e^{- H_{m=m_0}/T} e^{-(m_{HT}^2-m_0^2) V \avphi/2T}
	\; \; \delta \left( \avphi - \avphio \right) \, ,
\end{equation}
but we can now use the delta function to replace $\avphi$ in the 
exponential with $\avphio$, which is not integrated over; pulling it out
of the integral gives
\begin{eqnarray}
\frac{F( \avphio )}{T} & = & - \ln e^{-(m_{HT}^2-m_0^2) V \avphio / 2T}
	\int {\cal D} (A_i,\Phi) e^{-H_{m=m_0}/T} 
	\; \; \delta \left( \avphi - \avphio \right) \nonumber \\
& = & \left[ \frac{(m_{HT}^2-m_0^2) V}{2T} \avphio \right]
	- \ln \int {\cal D} (A_i,\Phi) e^{-H_{m=m_0}/T} 
	\; \; \delta \left( \avphi - \avphio \right) \, .
\la{mH_pulls_out}
\end{eqnarray}
The second term here is independent of $m_{HT}^2$; it can be determined
once and used at any value of $m_{HT}^2$ thereafter.
Hence the effect on $F(\avphi)$ of shifting $m_{HT}^2$ is very simple;
it just adds an extensive, linear in $\avphi$ term to $F$.  (The same
thing would happen if we considered the magnetization in a ferromagnetic
system where we vary the external magnetic field, or the
density in a liquid-gas system where we vary the pressure.  The key is
to consider a measurable which appears in the Hamiltonian next to the
control parameter which takes us through the transition.)  If we used
a different measurable the qualitative behavior would be the
same--the free energy as a function of that measurable  
would be roughly a ``tilted'' version of its $T_{\rm eq}$
appearance--but this would not hold as an exact quantitative statement.

\begin{figure}[tbh]
\centerline{\epsfxsize=6.2in\epsfbox{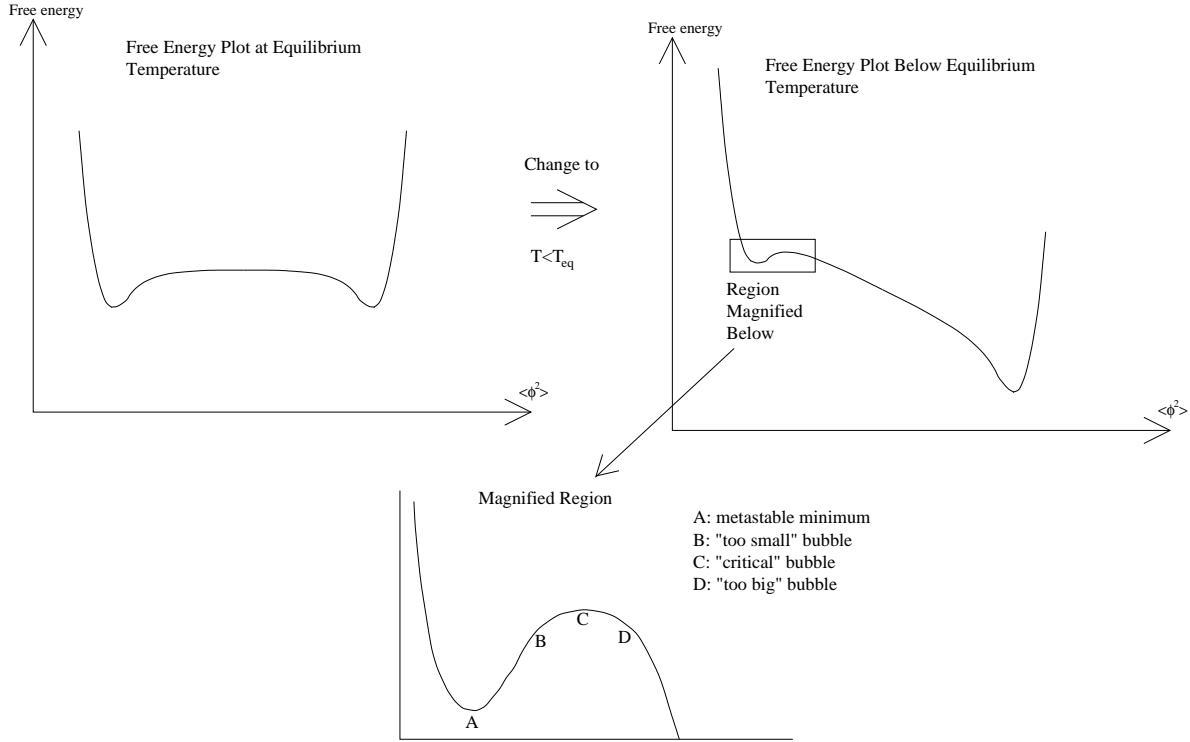}}
\vspace{0.25in}
\caption{\la{maxwell3} Cartoon showing how free energy in a finite
box changes when we lower the temperature.  For small temperature
changes, both minima survive, but one 
(A) is no longer globally stable.  The least likely configuration on the 
way to the stable minimum is (C) the critical bubble:  its unlikelihood
restrains the rate at which configurations near (A) go to the true
minimum.}
\end{figure}

Now consider how the free energy plot looks when we shift $m_{HT}^2$.  A 
cartoon is provided in Fig.~\ref{maxwell3}.  For small amounts of
supercooling, the symmetric phase minimum shifts over slightly, but persists
as a local minimum of the free energy.  It is labeled $(A)$ in the
figure.  Since we will be concerned with classical dynamics, for an element
of the thermal ensemble at $(A)$ to get to the global minimum, it must
pass through $(B)$, $(C)$, and $(D)$.  
The time evolution of a configuration at $(B)$ is almost certain, in the
medium term, not to go to $(C)$, because $F/T$ is $- \log({\rm
Probability})$; there are vastly more states with $\avphi$ equal the
value at $(B)$ than the value at $(C)$, so only a
tiny fraction will evolve to $(C)$ in the medium term, since time
evolution preserves the canonical ensemble.  On the other hand,
configurations at $(D)$ are almost certain not to ``go back,'' and will
continue to the broken phase minimum.  It is the rarity of
configurations at $(C)$ which limits the rate at which configurations
near $(A)$ time evolve into the broken phase.  If it were true that
every configuration at $(C)$ were on the way from $(A)$ to the broken
phase (or going from the broken phase to $(A)$), 
then with a little dynamical
information we could determine the nucleation rate.  It may be, though,
that most configurations at $(C)$ are either coming from $(B)$ and going 
back there, or coming from $(D)$ and going back there.  (This might
happen if the choice of measurable is not optimal, for instance.)
In this case, nucleations from the symmetric to the broken
phase would be even rarer than the free energy of states at $(C)$
implies.  So the rarity of 
configurations at $(C)$ provides an upper bound on the nucleation rate,
which can be turned into a determination with some additional dynamical
information.  

\subsection{Real time rate, dynamical prefactor}
\la{real_time}

Now we discuss how to turn the discussion and
cartoons of the last section into a calculation of a real time rate for
nucleations.  Suppose we have, by multicanonical tools discussed in
subsection \ref{multi_subsec},
computed the constrained free energy as a function of some
measurable, which we take to be $\avphi$.  It is also possible for us to
collect a sample of the canonical ensemble restricted to some narrow
range of $\avphi$, for instance, the range right around $\avphiC$, the
least likely value of $\avphi$.  What to we do with them?

The first thing to do is to determine how exponentially suppressed
critical bubbles are.  But the answer depends on the measurable and 
does not give a real time rate.  The second step is to determine what
the flux of states in the canonical ensemble through the critical bubble 
is, normalized to the probability to be in the symmetric phase.  That
is, we should determine
\begin{equation}
\la{flux_eq}
{\rm probability \; flux} \equiv \frac{ 
	{\rm P}(|\avphi - \avphiC| < \epsilon/2)}
	{ \epsilon \: {\rm P}(\avphi < \avphiC)}
	\times \left\langle \left| \frac{\Delta \avphi}{\Delta t}
	\right|_{\avphiC} \right\rangle \, ,
\end{equation}
with ${\rm P}({\rm condition})$ denoting the fraction of the canonical
ensemble which satisfies the condition, and 
with $\epsilon$ infinitesimal.  The first term here is the probability
density to be at the critical value $\avphiC$ of $\avphi$, which
is our ``definition'' of the critical bubble.  It could equally be
written as
\begin{equation}
\la{flux_2}
\frac{ {\rm P}(|\avphi - \avphiC| < \epsilon/2)}
	{ \epsilon \: {\rm P}(\avphi < \avphiC)}
	= 
        \left[ \int_{\rm small}^{\avphiC}
	\exp \left( \frac{F(\avphiC) - F(\avphi)}{T} \right)
	d \avphi \right]^{-1}\, ,
\end{equation}
and will be evaluated by Monte Carlo.
The second term in Eq.~(\ref{flux_eq}) is the mean change, in absolute
magnitude, of $\avphi$ in a time interval $\Delta t$, sampled over
configurations at the critical bubble.  Provided we take $\Delta t$
shorter than any typical infrared scale --- in particular it should be
shorter than the time scale to go from a configuration with $\avphi =
\avphiC$ to one where $F/T$ differs by order 1 --- then the ratio
will tell the number of configurations which pass from one side of the
critical bubble to the other in a time interval $\Delta t$, times the
time interval and divided by the number of symmetric phase
configurations.  In other words, the combination gives the flux
of configurations in the canonical ensemble through the critical bubble.

This flux is {\em NOT} the bubble nucleation rate we are after, though
it is clearly an upper bound.  
We have to multiply by the fraction of critical bubble
crossings which actually mediate a change from phase to phase.
To this end we define a ``dynamical prefactor,'' $\d$, as
\begin{equation}
\d \equiv \frac{{\rm trajectories \;
getting \; from \;}(B) \; {\rm to} \; (D)}{{\rm crossings \; of
}\; (C)} \, .\la{ddef}
\end{equation}
In other words, if we consider all real time trajectories, $\d$ is the
fraction of crossings of $(C)$ which represent ``permanent changes''
from one side to the other of the barrier.  To determine it we sample
the ensemble of configurations restricted to those at the critical
bubble, and for each element of this ensemble we construct a trajectory
forward and backwards in time, long enough to see the configuration
come from and go to an exponentially more common value of $\avphi$.  
Sampling trajectories every $\Delta t$, $\d$ is
\begin{equation}
\la{eq_for_d}
\d = \left\langle
	\frac{1}{\# {\rm \; of \; crossings}} \times 
	\left\{ \begin{array}{ll} 1 \quad  & {\rm change \; sides} \\
	0 & {\rm don't} \\ \end{array} \right. \right\rangle
\end{equation}
The average is over the canonical ensemble, restricted to configurations
with $| \avphi - \avphiC | < \epsilon/2$, and over
trajectories through those configurations.  The measure to be used is
the canonical one, 
times $|\Delta \avphi / \Delta t|$ evaluated where the crossing
takes place; so the sample is precisely the sample of the flux of states
through $\avphi = \avphiC$.
In Eq. (\ref{eq_for_d}), (\# of crossings) means the number of crossings
of the critical bubble a trajectory makes in the medium term.
To determine this we need to follow the trajectory until it reaches an
exponentially more common value of $\avphi$, ie.~point $(B)$ or $(D)$ in
Fig.~\ref{maxwell3}.\footnote{We must also
check that this criterion is sufficient to ensure that it is very
unlikely for the trajectory to return again to $(C)$, which is not
necessarily ensured; the ensemble of 
configurations at $(B)$ which have just evolved from configurations at
$(C)$ is not the same as the ensemble of all configurations at $(B)$.
In practice this does not prove to be a problem.}  The trajectory
``changes sides'' if it goes from $(B)$ to $(D)$ or vice versa, and does 
not change sides if it returns to the same side it came from.
In our case, the real time evolution will be Langevin evolution, see
next section, and the forward and backwards time
evolutions are just two Langevin evolutions with two different realizations
of the random force.  For the case of
Hamiltonian dynamics, 
the forward evolution would be 
evolution with a set of momenta drawn from the canonical ensemble and
the backwards evolution would be evolution with the same momenta, but
sign reversed (T conjugated).  In either case we are approximating the
expectation values in Eq. (\ref{eq_for_d}) with an average over a sample
of trajectories, ie.~we take the average in Eq. (\ref{eq_for_d}) by a
Monte Carlo integration.  We discuss why this procedure gives the
correct nucleation rate at more length in Appendix \ref{separatrix}.

For Hamiltonian dynamics, in an UV regulated theory, 
both $\d$ and $\avchange$ should have
well defined small $\Delta t$ limits.  This is not the case for Langevin
dynamics, however.  If we sample a Langevin trajectory with a 
smaller $\Delta t$, the number of crossing should grow, as each crossing 
gets ``resolved'' into potentially more; this is a normal feature of a
Brownian path.  However, $\avchange$
will also depend on $\Delta t$ by a compensating amount.  For suitably
short $\Delta t$, the time history of $\avphi$ near each crossing looks
like a Brownian random walk.  By well known properties of Brownian
random walks, $\avchange$ scales as
$(\Delta t)^{-1/2}$, while $\d$ scales as $(\Delta t)^{1/2}$, and
the product has a finite small $\Delta t$ limit.  Hence, for Langevin
dynamics, neither $\d$ nor $\avchange$ are well defined but the
product is. 

It is finally the product,
\begin{equation}
\frac{{\rm nucleation \; rate\:}}{{\rm Volume}} 
= \frac{1}{2V} \: {\rm probability \; flux} \: \times \: \d \, ,
\end{equation}
which we are interested in.  The factor of $(1/2)$ is because half of
the permanent crossings the algorithm finds are {\em into} the symmetric 
phase.  The factor $1/V$ turns the nucleation rate into a rate per unit
volume.

\subsection{Complications:  peculiar behavior in finite volumes}
\la{finiteV}

In this subsection we discuss some complications with applying our
technique, arising from finite volume effects.  The conclusion will be
that the volume must be fairly large, so the bubble interior fills at
most about $15\%$ of it; and that as a consequence, it is best to choose
a measurable with a very small variance in the metastable phase.  The
impatient reader may want to skip this section and just accept that
conclusion.  

What we want is the nucleation rate in very large volumes.
In practice it is impossible to work directly in a very large volume,
for reasons of numerical cost.  Naively the cost of performing the
Monte Carlo calculation scales as the volume, but in practice the
scaling is still more severe, because of memory and communication costs
and because the Monte Carlo becomes less efficient in large volumes,
particularly at the value of $\avphi$ where the typical configuration
changes from being homogeneous, supercooled symmetric phase to being an
isolated bubble.

\paragraph{Resolving the critical bubble:}

There is another reason why we would like to work in a small volume.
This is because any volume averaged measurable, for example $\avphi$,
fluctuates also in the pure symmetric or broken phase.  The width of
these fluctuations behave as $1/\sqrt{V}$, $V$ the volume, whereas the
contribution of a fixed size bubble to $\avphi$ scales as $1/V$.  Thus,
if we keep the size of the critical bubble constant but increase $V$,
the fluctuations of $\avphi$ in the pure symmetric phase outside of the
bubble increase and degrade the the ``cleanliness'' of $\avphi$ as a
description of the critical bubble.

Concretely, what we mean by this degradation is that the more
symmetric phase there is, the more likely it is that the value of
$\avphi$ consistent with the critical bubble (point C in
\fig\ref{maxwell3}) really arises from too small a bubble plus an
upward fluctuation in the symmetric phase contribution to $\avphi$, or
too big a bubble and a downward fluctuation. 
The result is that, as we increase the volume, 
the {\em measured} free energy of the
bubble should fall somewhat faster than the 
``true'' bubble free energy (which 
decreases as $-\log(V)$, due to the translational 
zero modes of the bubble configuration).
However, the measured value of 
the ``dynamical prefactor'' $\d$, Eq.~(\ref{eq_for_d}),
should become smaller, since it is more likely
that an imputed critical bubble is really on one or the other side and
will begin and end at the same phase.  These effects cancel exactly, so that
the nucleation rate has a well-defined infinite volume limit.
However, as $\d$ becomes smaller, it takes more
work to measure it with good relative accuracy.  This effect also
degrades the efficiency of the Monte Carlo.  

\paragraph{Maximum size of the critical bubble:}

For the above reasons it is to our advantage to use as small a volume
for a given size bubble as we can get away with.  Naively, this 
means we should
use the smallest volume for which the critical bubble is unable to
``see itself'' around our periodic boundary conditions (we choose to
work in a cubic box with periodic boundary conditions.  It appears
necessary to use a box which has everywhere a flat spatial metric, and
does not have boundaries, because either effect could modify the bubble
free energy.).  Very naively,
this means we must ensure that the radius of the critical bubble
$r_{\rm max}$ is less than half the box length $L$, $r_{\rm max} <
L/2$.

\begin{table}[tb]
\centerline{\begin{tabular}{|c|c|c|c|} \hline
 Geometry  &  Sphere  & 
$\qquad$ Cylinder $\qquad$ &$\qquad$  Planes $\qquad$  \\ \hline
Area($r$) & $ 4 \pi r^2 $ & $ 2 \pi r L $ & $ 2 L^2 $ \\ \hline
$\qquad$ Volume($r$) $\qquad$ 
& $ \frac{4 \pi}{3} r^3 $ & $ \pi r^2 L $ & any \\ \hline
$A(V)$ & $\qquad$ $(36 \pi V^2)^{1/3}$ $\qquad$ 
& $2 \sqrt{\pi L V}$ & $2L^2$ \\ \hline
Volumes where stable & $0 \; - \; 4\pi/81$ & 
$4\pi/81 \; - \; 1/\pi$ & $1/\pi \; - \; (1 \! \! - \! \! 1/\pi)$ \\ \hline
Vol. where metastable & $0 \; - \; \pi/6$ &
 $1/4\pi \; - \; \pi/4$ & $0 \; - \; 1$ \\ \hline
\end{tabular}}
\vspace{0.3in}
\caption{\la{stability} Area and volume as functions of $r$, and area 
as function of volume, for each possible geometry for phase
coexistence; and derived volume range where the geometry is preferred
and where it is metastable in the strict thin wall limit.  The upper end 
of the sphere and cylinder metastability ranges are where they
touch themselves across the periodic boundary conditions.  The lower
metastability limit on the cylinder is where it becomes unstable to
a $\delta r = \sin(2\pi z/L)$ excitation.} 
\end{table}

\begin{figure}[tb]
\centerline{\epsfxsize=4in\epsfbox{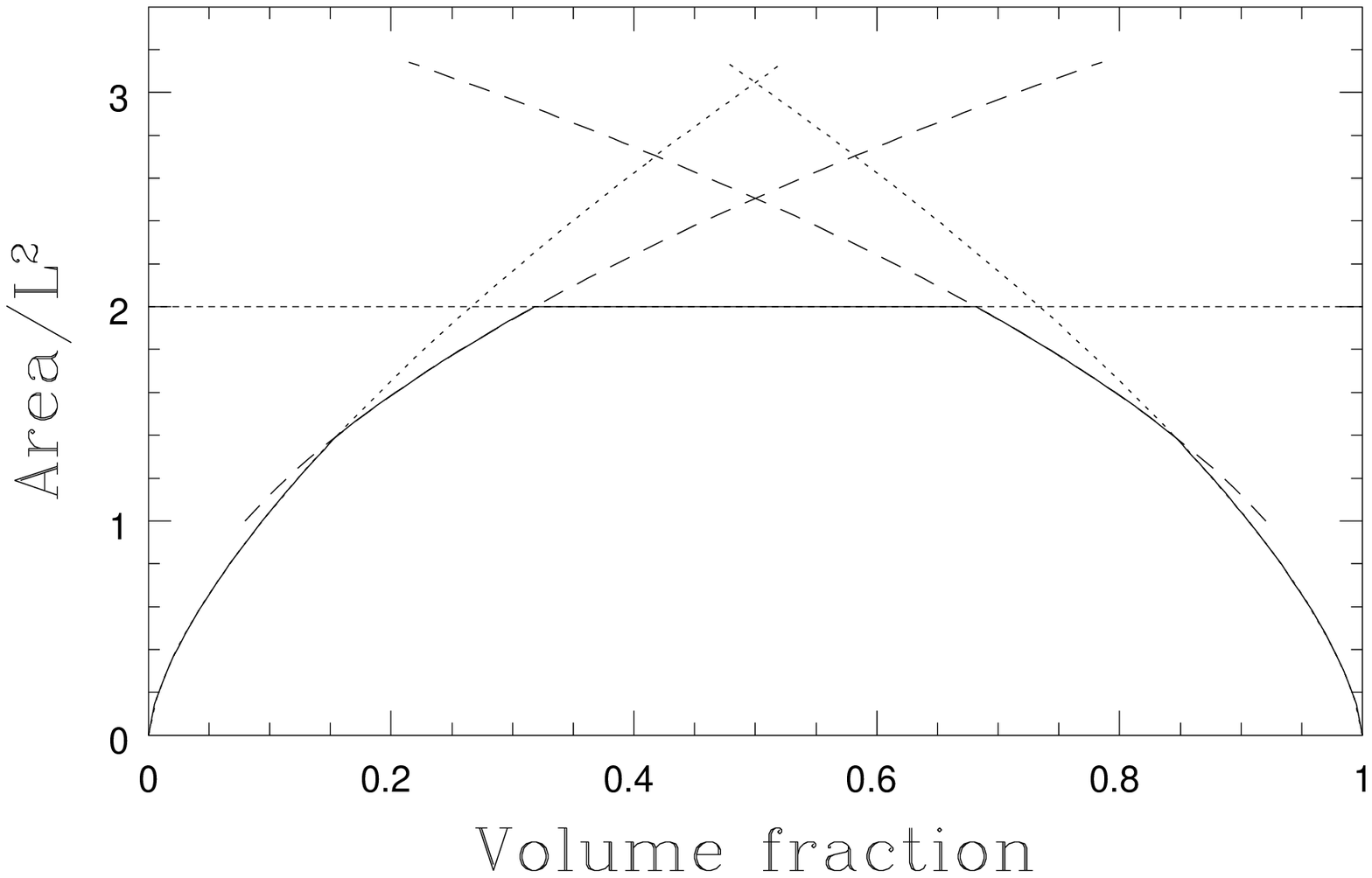}}
\vspace{0.2in}
\caption{\la{simple} Area, and hence free energy, as a function of
volume fraction, in the thin wall approximation.  
The solid line is the minimum over interface geometries; the
large volume free energy curve would follow the solid line.  Dotted
lines are the metastable extensions of the sphere geometry (sloping) or
the planar boundary geometry (flat), while the dashed lines show
metastable extensions of the cylindrical geometry.}
\end{figure}

In fact the criterion for a sufficiently large $L$ is more severe than
$L > 2r_{\rm max}$.  The reason is that a spherical broken phase bubble
with $L/3 < r < L/2$ is at best only metastable:
%bubble geometry, considered
%among all bubble geometries with the same broken phase volume fraction;
a configuration with a cylinder of the broken phase, extending through the
length of the box and having the same volume as the bubble, will have smaller
phase interface area and hence smaller total free energy.
To compare the favorability of
different geometries as a function of broken phase volume fraction, we
will make a ``thin wall'' approximation in which the interface between
phases is treated as a geometrical surface, and the free energy is
equal to its area times the surface tension: $F=\sigma A$.
This approximation is correct in the limit
that the box size $L$ is much larger than the wall thickness.  Though
our simulations will not be strictly in this limit, it is a suitable
approximation for understanding the relative favorabilities of different
mixed phase geometries.  
% In this approximation, the free energy is
% proportional to the interface area, $F = \sigma A$.
One can then write down how the area and volume vary as a function of
radius for a sphere and a cylinder extending the length of the box, and
for a pair of planar interfaces.  The results are shown in Table
\ref{stability} and compared in Figure \ref{simple}.  Equating the areas 
at fixed volume, 
\begin{equation}
A(V)_{\rm sphere} = (36 \pi V^2)^{1/3} = \sqrt{4 \pi L V} =
A(V)_{\rm cylinder} \, ,
\end{equation}
we find that the sphere and cylinder are degenerate for 
$V = L^3 \times 4 \pi / 81 \simeq 0.155 L^3$, and the cylinder and planes 
are degenerate for $V = L^3 / \pi$.  A plot of the free energy 
in a very large box should approximately resemble the solid line in 
Fig.~\ref{simple}.  It will have slope
discontinuities where one geometry gives way to the next.  The
metastability of one geometry against another can be quite strong, and
this makes it challenging to
perform a multicanonical Monte Carlo determination of the free energy 
in a volume large enough that the
interfaces are thin compared to $L$, unless one only wants the plot in
the region where one geometry is relevant.  (Fortunately this will be
the case for us.)

Since both the cylinder and the
planar geometry have free energies which clearly depend essentially on
the box geometry and volume, the part of the free energy plot where they 
dominate describes finite volume artifacts rather than physics which has 
any correspondence to that at larger volumes.  However, the radius of
the sphere where the cylinder becomes equally favorable is only $r =
L/3$, safely small enough that the sphere will not ``see itself'' around 
the periodic boundary conditions (unless $L$ is only a few times the
interface width).  Hence we can expect the sphere geometry, until it
stops being the favored geometry, not to care about the periodic boundary 
conditions but to represent more or less faithfully the behavior of an
isolated bubble in a larger volume.

We may also expect that, in a suitably large volume, the transition
rate between the spherical and cylindrical geometries is sufficiently
slow that we could rely on the metastability of the sphere to explore
larger radii than $r=L/3$.  However we will conservatively not do so
in what follows, but will restrict ourselves to such volume and bubble
size combinations that the critical bubble we obtain will have $\avphi
< 0.15 \avphi({\rm broken}) + 0.85 \avphi({\rm symmetric})$.  We will
also check to see that the bubbles we analyze are approximately
spherical and not cylindrical and that they do not touch across the
periodic boundary.

%------------------------------------------------------------------
\section{Electroweak Bubble Nucleation as a Problem in Classical
Statistical Mechanics}

\la{statmech}

In this section we briefly review why we can view the bubble nucleation
problem as a problem in classical statistical mechanics for Yang-Mills
Higgs field theory.  Nothing in this section is new; it reviews
the last few years' developments both in the thermodynamics of the
electroweak phase transition and in the dynamics of infrared Yang-Mills
Higgs fields.  We include it here to make the paper more
self-contained.  Readers who are already familiar with this material may 
want to skip this section.

\subsection{Thermodynamics:  dimensional reduction}

Here we review how the thermodynamics of infrared fields in the SU(2)
sector of the standard model is well approximated by a 3 dimensional
path integral, which is the same as the partition function of classical
3+1 dimensional SU(2) Higgs theory at finite temperature (and with
suitable regulation and counterterms).

The full theory we are interested in has thermodynamics described by the 
path integral (Lorentz indices are Euclidean with positive metric; Greek 
indices range over space and time, Latin indices $i,j,k$ only over
space, indices $a,b,c$ are SU(2) group indices)
\begin{eqnarray}
Z & = & \int {\cal D}(\Phi , A_\mu , {\rm etc}) 
	\exp ( - S_{\rm E} / \hbar ) \, , \nonumber \\
S_{\rm E} & = & \int_0^{\hbar/T} d\tau \int d^3 x \bigg[ 
	\frac{1}{4 g^2} F_{\mu \nu}^a F^{\mu \nu}_a + 
	(D_\mu \Phi)^\dagger (D^\mu \Phi) 
	+ ( m_{Ho}^2 + \lambda \Phi^\dagger \Phi ) \Phi^\dagger \Phi
\la{action}	\\
& & \hspace{1.2in} + {\rm \: hypercharge \: + \: fermions \: + \: glue \;}
	\bigg] \, . \nonumber 
\end{eqnarray}
Here and throughout $g$ is $g_w$ the weak coupling.  The $\tau$
integration has periodic boundary conditions for bosons and antiperiodic 
boundary conditions for fermions.  Beginning here we will neglect
hypercharge, to simplify things.  This is not too bad an approximation
\cite{KLRS_U1}.  Including it would be a straightforward extension
of what we discuss.

There are two things to observe right away about this theory.  First,
mean field theory predicts that if $m_{Ho}^2$ changes, there is a second
order phase transition.  Second, at the scale $T$
(which will be of order the weak scale, $T \sim 80$GeV), the coupling
is weak.  This is just the statement that the weak sector of the
standard model is indeed weakly coupled.
Hence, if there is a phase transition, barring some large
hierarchy of couplings such as $\lambda / g^2 \ll 1$, it will be weak,
and correlation lengths will be $\xi \gg 1/T$.

\begin{figure}[tbh]
\centerline{\epsfxsize=5in\epsfbox{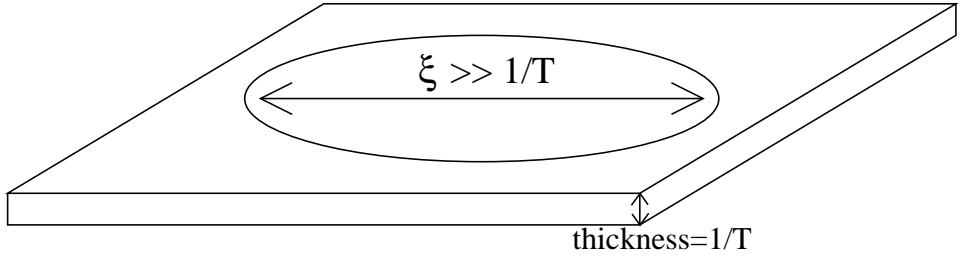}}
\vspace{0.25in}
\caption{Cartoon of how a 3+1 dimensional spacetime, $\Re^3 \times S^1$,
drawn here as 2+1 dimensional,
can look effectively 3 dimensional for long distances. \la{4to3}}
\end{figure}

As motivated in Fig.~\ref{4to3}, at such infrared scales the effective
behavior is 3 dimensional.  This is just because any field varying only
on the length scale $\xi \gg 1/T$ will not vary appreciably across the
Euclidean time width of the ``slab.''  One sees this formally by Fourier 
transforming the $\tau$ direction in Eq.~(\ref{action}).  The Euclidean
frequencies arise from transforming a compact range and so are discrete: 
$(\partial_\tau \Phi(\tau))^2$ becomes $(2\pi n T)^2 \Phi_n^2$ for bosons,
while $\overline{\psi} \gamma^0 \partial_\tau \psi(\tau)$ becomes
$((2n+1) \pi T) \overline{\psi}_n \gamma_0 \psi_n$ for fermions.  All but
the $n=0$ bosonic mode are very heavy and can be integrated out.  The
result (continuing to use 4 dimensional notation for fields and
couplings) is
\begin{eqnarray}
\frac{S_{\rm E}}{\hbar} & = & \frac{1}{T} \int d^3x \bigg[ 
	\frac{1}{4g^2} F_{ij}^a F^{ij}_a 
	+ \frac{1}{2} (D_i A_0)^a (D_i A_0)^a 
	+ (D_i \Phi)^\dagger (D_i \Phi)
	+ m_{HT}^2 \Phi^\dagger \Phi
	+ \lambda ( \Phi^\dagger \Phi )^2
	\nonumber \\ & & \hspace{0.6in}
	+m_{\rm D}^2 A_0^a A_0^a 
	+ \frac{(g^2 + O(g^2 \alpha_w))}{4} 
	A_0^a A_0^a \Phi^\dagger \Phi
	+ O(g^2 \alpha_w) (A_0^a A_0^a)^2
	+ { \rm \: Dim \; 6 \:} \bigg] \, ,  
\la{S_dimreg}
\end{eqnarray}
where ``Dim 6'' indicates dimension 6 and higher induced operators,
which have an irrelevantly small effect on the infrared physics, so we
immediately drop them.
The fields we write here correspond to the zero frequency components of
the 4 dimensional fields, ie.~$\Phi$ in Eq.~(\ref{S_dimreg}) is 
$T \int d\tau \Phi(x,\tau)$, with $\Phi(x,\tau)$ the field appearing in
the path integral in Eq.~(\ref{action}).  In
this expression $\lambda$ and $g^2$ are couplings of an effective $3D$
theory, which (after dropping dimension 6 operators)
is super-renormalizable; they do not run.  Their relation 
to the coefficients of the full theory, and a detailed discussion of the 
matching procedure used to derive them, is given in \cite{KLRS}.  In
particular we mention that the Higgs mass squared, $m_{HT}^2$, receives
$O(g^2 T^2)$ positive thermal corrections;
\begin{equation}
\la{m_of_T}
m_{HT}^2 = m_{\rm vac,ren}^2 + \frac{3 g^2 + 4 y_t^2 + 8 \lambda}
	{16} T^2 + (\: \mu {\rm \; dependent \;} O(g^4 T^2) ) \, .
\end{equation}
Here $y_t$ is the top quark Yukawa coupling.  Since $m_{\rm vac,ren}$ is 
negative, varying $T$ can change the sign of the Higgs mass squared and
induce a phase transition.\footnote{Note that both $m_{HT}^2$ and
$m_{\rm D}^2 \sim g^2 T^2$ renormalize logarithmically at the two loop
level, $(\partial m_{HT}^2/\partial \ln\mu) \sim g^4 T^2$, 
so the sign of $m_{HT}^2$ near the transition is actually
renormalization point dependent.}  For
generic values of $\lambda / g^2$, correlation lengths at the phase
transition are of order $\xi \sim 1 / g^2T$, which is why we can drop
the dimension 6 operators.

Note that the form of $S_{\rm E}/\hbar$ looks very much like the
Boltzmann factor, $E/T$, of a
classical theory.  The main difference is that in classical SU(2) Higgs
theory, we expect the electric field strength and the Higgs field
momentum $\Pi$, rather than $A_0$, to
appear;
\begin{eqnarray}
Z_{\rm cl} & = & \int {\cal D}(E_i^a , A_i^a , \Phi , \Pi) \exp(-H/T) 
	\; \delta \! \left( (D_i E_i)^a 
	+  \left[ \Phi^\dagger \frac{i g \tau^a}{2} \Pi 
	+ {\rm c.c.} \right] \right)
	\, , \nonumber \\
H & = & \int d^3x \bigg[ \frac{1}{4g^2} F_{ij}^a F^{ij}_a 
	+ \frac{1}{2} E_i^a E_i^a
	+ \Pi^\dagger \Pi
	+ (D_i \Phi)^\dagger (D_i \Phi)
	+ m_{HT}^2 \Phi^\dagger \Phi
	+ \lambda ( \Phi^\dagger \Phi )^2 \bigg] \, .
\la{H_2}
\end{eqnarray}

Here the delta function enforces Gauss' Law.  But as noted by Ambj{\o}rn 
and Krasnitz \cite{AmbKras}, if we implement Gauss' Law by introducing a 
Lagrange multiplier, which we suggestively name $A_0$,
\begin{equation}
\delta \left( \ldots \right)
	= \int {\cal D} A_0 \exp \left( i \int d^3 x A_0^a 
	\left( (D_i E_i)^a +
	\left[ \Phi^\dagger \frac{i g \tau^a}{2} \Pi 
	+ {\rm c.c.} \right] \right)
	/ T \right) \, ,
\end{equation}
then the $E$ and $\Pi$ integrations are Gaussian and can be performed,
generating 
\begin{equation}
\la{H_3}
H \supset \frac{1}{2} (D_i A_0)^a (D_i A_0)^a
	+ \frac{g^2}{4} A_0^a A_0^a \Phi^\dagger \Phi
	+ (0 + {\rm \: radiatively \; induced \:}) A_0^a A_0^a \, .
\end{equation}
Here the bare Debye mass is zero but one is radiatively induced, with a
linear divergent coefficient in any regulation where such terms do not
identically vanish (such as the lattice).  Hence, the thermodynamics of
the full theory DO look like those of the classical theory, except that
the Debye mass of the classical theory is radiatively induced and
regulation dependent, and there are very small $O(g^2 \alpha_w)$
extra interaction terms involving the $A_0$ field, present in the actual 
thermodynamics but not in the thermodynamics of the classical system.

It is a good approximation, for the thermodynamics both of the classical 
theory and of the dimensionally reduced full theory, to integrate out
the $A_0$ field, including it by the radiative corrections it will
induce in the remaining couplings.  In this approximation one shifts
slightly the coefficients of the terms in Eq.~(\ref{S_dimreg}) which do
not contain $A_0$, and drop those which do.  The change in the
coefficients is computed in \cite{KLRS}.  After this approximation, the
classical thermodynamics and the dimensionally reduced thermodynamics
coincide exactly.  The final form of the partition function describing
the thermodynamics is then
\begin{eqnarray}
Z & = & \int {\cal D} ( A_i , \Phi ) \exp ( - H / T ) \, , \nonumber \\
H & = & \int d^3 x \bigg[ \frac{1}{4 g^2} F_{ij}^a F_{ij}^a + 
	(D_i \Phi)^\dagger (D_i \Phi) 
	+ m_{HT}^2 \Phi^\dagger \Phi
	+ \lambda (\Phi^\dagger \Phi)^2 \bigg] \, .
\la{3D-theory}
\end{eqnarray}
If this final integration over the $A_0$ field is not deemed reliable
enough it is straightforward to modify what we will do below to include
it in the thermodynamic calculation.

We will also mention a few results, perturbative and nonperturbative,
which have been obtained for the partition function shown above.
Perturbatively we can describe the strength of the phase transition by
studying the effective potential for the (gauge fixed) Higgs field $\phi$,
$\phi = \sqrt{2 \Phi_{\rm cond}^\dagger \Phi_{\rm cond}}$.  Note that
there is no good gauge invariant, nonperturbative way to separate the
condensate from the fluctuations.  When gauge field fluctuations become
large, the whole perturbative approach becomes questionable.
Nevertheless, when the phase transition is strong, perturbation theory
is useful, essentially because gauge field fluctuations are suppressed
in the broken phase (which is therefore well described).
At one loop, and neglecting scalar loops as is appropriate in the
$\lambda / g^2 \ll 1$ approximation, 
\begin{equation}
V_{\rm 1 \; loop} = \frac{m_{HT}^2}{2} \phi^2 
	- \frac{g^3}{16 \pi} \phi^3 T + \frac{\lambda}{4} \phi^4 \, .
\la{1loop}
\end{equation}
This effective potential can have two minima because of the (one
loop) negative $\phi^3$ term.  Since it is a loop effect which allows a
first order phase transition, we say the transition is radiatively
induced first order.  At the transition, 
the broken phase value of $\phi$ is such that the
cubic term and (tree level) $\lambda \phi^4$ term are of the same
order.  Since this requires a one loop effect to be of the same size as
a tree level one, it either implies that $\lambda / g^2 \ll 1$, or that
perturbation theory will not be a reliable expansion.  Hence
perturbation theory determines attributes of 
the transition at best as an expansion in $\lambda / g^2$.  It is this
relatively poor performance for perturbation theory which makes a
nonperturbative treatment necessary.

Nonperturbatively it is known that, as expected, the phase transition is
well described by perturbation theory for small $\lambda / g^2$; but
perturbation theory is completely wrong for larger values
\cite{KLRS_results,KLRS_cross}.   In fact there is NO phase
transition in the MSM above a critical value $\lambda / g^2 \simeq
0.0983 \pm 0.0015$ \cite{endpt}.  In extensions of
the standard model with new bosons which are light at the phase
transition, we must include the new light bosons in 
the effective theory considered.  At least for the case of an added
scalar top, the strength of the phase transition is significantly
enhanced \cite{BodekerLaineSchmidt,LaineRummukainen}.  It would be
straightforward but more numerically expensive to apply the tools
developed here to this physically interesting case.

\subsection{Dynamics:  classical effective theories}
\la{dyn_subsec}

The infrared thermodynamics of the SU(2) sector of the standard model 
match those of classical 3+1 dimensional SU(2) Higgs theory, as
discussed above.  Does this matching also apply at the dynamical level,
as originally conjectured (in a 1+1 dimensional context) by Grigoriev
and Rubakov \cite{GrigRub}?  In other words, are the dynamics of the
infrared SU(2) Higgs fields described by classical Hamiltonian dynamics?

The answer is ``no, it is more complicated than that.''\footnote{In
defense of Grigoriev and Rubakov we should mention that the
complications discussed in this subsection do not arise in 1+1
dimensions, where the UV behavior is much more mild; hence their
conjecture {\em is} correct for the problem they were addressing, namely 
the dynamics of the 1+1 dimensional abelian Higgs model.  The problem is
applying it to the 3+1 dimensional problem of interest instead.} 
As we now discuss, the dynamics of infrared gauge and Higgs fields are
indeed classical; but they are 
not described by classical Hamiltonian dynamics.  They are, to
leading order in the {\em logarithm} of $g$, described by ``classical''
Langevin dynamics.  Because a factor of 2 error in the treatment of the
dynamics will change our determined rate by $\sim \exp(\pm 1)$, while
the rate itself is $\sim \exp(-100) T^4$, we will find it sufficient to
make this approximation and take the dynamics to be Langevin.

To see that the dynamics of the real system are not classical
Hamiltonian dynamics, look again at the thermodynamic discussion,
especially Eq.~(\ref{H_2}) and Eq.~(\ref{H_3}).  We see that there are
linear UV divergences in the classical Debye mass for the $A_0$ field;
and yet the $A_0$ field is related to the electric fields $E$, which
generate the Hamiltonian dynamics for the gauge fields.  This implies
that there are divergent UV corrections to the gauge field dynamics of
the classical fields.  As first advocated by B\"{o}deker, McLerran, and
Smilga \cite{Smilga}, we should consider this a potential problem for
the study of the classical field dynamics.  
In fact, as argued by Arnold, Son,
and Yaffe \cite{ASY}, what it means is that the classical Hamiltonian
dynamics do not have a good regulation independent limit.  Hence, the
infrared gauge field dynamics of the classical theory technically do not 
exist.  The dynamics of the full theory can scarcely coincide with those 
of the classical theory if the classical theory's dynamics are sick.

The physical origin of this problem is actually well known plasma
physics.  Transverse electric fields in a plasma feel Landau damping.
This leads to very slow, overdamped evolution of infrared magnetic
fields, as is typical in a conducting medium.  As the classical theory
cutoff is lifted, there are more and more ``plasma'' degrees of freedom, 
and the damping becomes ever more efficient.  The correct treatment is
to make the damping have the same efficiency as in the quantum theory.
This requires studying the classical theory with hard thermal loop (HTL) 
effects \cite{HTL_papers} included.  (The hard thermal loops are the
nonabelian generalization of Debye screening, Landau damping, and other
plasma effects familiar from electromagnetic plasmas.)
Such a classical, but HTL
including, treatment should be correct at leading order in the coupling
$g$.  Two numerical implementations of
such a classical theory now exist; one \cite{particles} is based on a
proposal by Hu and M\"{u}ller \cite{HuMuller}, and one \cite{BMR} is based
on a proposal by B\"{o}deker, McLerran, and Smilga \cite{Smilga}, and
more recently discussed by Iancu \cite{Iancu_recent}.  Both are extremely
complicated.  Probably the second method could be utilized in the type of 
computation we are going to discuss, but the numerical effort would be
substantially greater than what we discuss below.

As first demonstrated by B\"{o}deker \cite{Bodeker}, and further
discussed and clarified both by B\"{o}deker \cite{moreBodeker},
Arnold, Son, and Yaffe \cite{moreASY}, and Litim and Manuel
\cite{LitimManuel}, inclusion of the hard thermal loops
in the infrared equations of motion is actually unnecessary, because at
leading order in $\log(1/g)$ the dynamics of the gauge fields are simple 
Langevin dynamics.  We will not attempt to reproduce their arguments in
detail here, but will only physically motivate them.  

First, consider
the behavior of an abelian plasma.  We will distinguish two
characteristic length scales:  the Debye length $l_{\rm D} \sim 1 /
m_{\rm D} \sim 1 / gT$, and the scattering length $l_{\rm scatt} \sim 1
/ g^4 T \log(1/g)$.  The former is the shortest length scale where plasma
effects are important.  The latter is the mean length 
over which a current can propagate, before it is disrupted by collisions 
in the plasma.  It coincides with the free path of a charge carrier to
undergo large angle scattering.

On scales much longer than $l_{\rm scatt}$, a magnetic field evolves as
if it were in a conductor: $D \times B = j = \sigmaE E = \sigmaE D_0 A$, 
where $\sigmaE$ is the electric conductivity.  
This is just Langevin dynamics; the time derivative of the field $D_0 A$ 
is proportional to $-dH/dA = D \times B$.  In a thermal bath there is
also a ``noise'' term uniquely determined by the thermodynamics
(fluctuation dissipation).  On scales between $l_{\rm D}$ and $l_{\rm
scatt}$ the story is significantly more complicated because the
homogeneity scale is shorter than the
free path over which an electric current propagates.  In this regime the
field evolution is described by an equation with a wave
number dependent conductivity, which is nonlocal in real space.

Now consider the case of a nonabelian theory.
The key difference is that, for the nonabelian theory, $l_{\rm scatt}
\sim 1 / g^2 T \log(1/g)$.  The reason is that nonabelian collisions
{\em exchange nonabelian charge} (``color''), 
so {\em any} collision, however soft, can destroy
the current a particle is carrying.  Hence $l_{\rm scatt}$ is of order
the mean free path for {\em any} scattering, not just large angle
scattering.  Hence, on the length scale $1/g^2 T$, the dynamics are
given by a simple Langevin equation \cite{AY2_long},
\begin{equation}
\sigmaE D_0 A_i = - \frac{\partial H}{\partial A_i} \, , \qquad
\sigmaE^{-1} = \frac{3}{m_D^2}  \gamma \, , \qquad
	\gamma = \frac{N_{\rm c} g^2T}{4\pi} 
	\left[ \ln \frac{m_{\rm D}}{\gamma} + 3.041 \right] \, ,
\la{Bod_th}
\end{equation}
with $N_c=2$ for our SU(2) application.  Here $\sigmaE$ is the
nonabelian (``color'') conductivity.

The extension of these ideas to the case where there is a Higgs field
turns out to be remarkably simple; one also evolves the Higgs fields
under Langevin dynamics, but giving the Higgs fields a much larger
diffusion constant than the gauge fields.  For a discussion see
\cite{Bodek_higgs}.  These are the dynamics we will apply for the real
time part of our studies below.  Note that they are only justified at
(next to \cite{AY2_long}) leading order in $1/\log(1/g)$, not a very good
expansion; but we are willing to accept an approximation which will
yield an error of $\pm 1$ in the exponent of the nucleation rate, since
the rate itself is $\sim \exp(-100)$.

%------------------------------------------------------------------
\section{Computational Details and Numerics}
\la{Numerics}

\subsection{Our choice for a measurable}
\la{variance}

In the discussion above we always chose to consider $\avphi$, the
space averaged Higgs field length squared, as the observable used to
distinguish the phases and the critical bubble.  This has been the
traditional ``order parameter'' observable in Monte Carlo simulations
of SU(2) + Higgs theory.  It is easy to measure, and, because its
variance is UV finite, given large enough volume, it can unambiguously
separate the symmetric and broken phases.  Also, as we have seen, it
is an extremely convenient choice because it makes it quite easy to
use one set of multicanonical data to study a range of temperatures.
Further, after UV counterterm subtractions, it has a good zero lattice
spacing limit.

However, {\it a priori} it is not obvious that this (or almost any
other) measurable can distinguish the critical bubble well enough
for practical calculations.  In particular we might worry that there is
too large a ``noise'' contribution from the $85\%$ of the volume which
must be in the symmetric phase (see the discussion 
in subsection \ref{finiteV}).
A necessary (but insufficient) criterion for a good measurable, needed
to avoid this problem, is
that it has a small variance in the symmetric phase.  
It turns out that $\avphi$ does
indeed have a very small variance in the symmetric phase.  The leading order
perturbative result for the variance of $\avphi$ in the symmetric phase, 
in a volume $V$ much larger than $1/m_{\rm symm}^3$, with $m_{\rm symm}$ 
the symmetric phase scalar mass, is
\begin{eqnarray}
\sigma_{\avphi,\, {\rm symm}}^2 
	& = & \frac{1}{V^2} \int \! d^3 x \, d^3 y \, \langle \phi^2(x)
	\phi^2(y) \rangle_{\rm connected} \nonumber \\
	& = & \frac{4}{V} \int \frac{d^3 p}{(2\pi)^3} \,
	\frac{T^2}{(p^2 + m^2)^2} = \frac{T^2}{2 \pi m V}
	\simeq \frac{4T \sqrt{2 \lambda / g^2}}{g^2 V} \, ,
\end{eqnarray}
where in the last approximate equality we have substituted in the
equilibrium, one loop symmetric phase Higgs mass for $m$.
This variance is to be compared to the broken phase variance, which gets 
an added contribution from fluctuations in the zero mode, which has a
condensate $\phi_0$:
\begin{equation}
\sigma_{\avphi, \,{\rm broken}}^2 = \frac{2 \phi_0^2 T}{m^2 V} +
	\sigma_{\avphi,\,{\rm symm}}^2 = \frac{4 T}{(\lambda / g^2)
	g^2 V} + \sigma_{\avphi,\,{\rm symm}}^2 \, ,
\end{equation}
which is much larger for small $(\lambda /g^2)$.  To get a phase
transition strong enough to preserve baryon number after its conclusion, 
we will consider the case $\lambda / g^2 = 0.036$; for this case the
broken phase variance is about 100 times larger, and if $15\%$ of
the volume is in the broken phase, the variance contributed by this
volume greatly exceeds that contributed by the symmetric phase; symmetric
phase fluctuations will not pose a problem.  Also note that the broken
phase fluctuations are dominated by the motion of the condensate; and we 
expect that fluctuations in the condensate size are directly important
to whether a bubble is more or less than critical, so these fluctuations 
may also not be dangerous.

{\it A posteriori} we will of course determine whether the choice of
measurable was a good one.  For instance, when determining ${\bf
d}$, we can determine what fraction of trajectories crossing $\avphi =
\avphiC$, actually lead to a nucleation.  We will find that $\avphi$
is in fact a good measurable; the fraction of trajectories crossing the
critical bubble which lead to a nucleation
is statistically compatible with 1/2, which is the maximum
possible under Langevin dynamics.  Note however that the
above arguments suggest that, if we were studying nucleation out of the
{\em broken} phase, then $\avphi$ would probably be a very bad
measurable.  We will not discuss what measurables might be useful for
studying this nucleation rate, which fortunately is not cosmologically
interesting.

Note that the value for the scalar self-coupling we use, produces a
Higgs mass lighter than the experimental limit; in fact there is {\em
no} physical Higgs mass which gives such a small value for the ratio of
$\lambda / g^2$ (parameters of a 3-D effective theory) \cite{KLRS}.  We
study this case as a toy example, because the phase transition is
relatively strong here and perturbation theory arguably should be
reasonable. 

\subsection{Multicanonical method}
\la{multi_subsec}

As described in Sect.~\ref{Plan}, by far the dominant factor in
the bubble nucleation rate is determined by the constrained free
energy $ F(\avphi) = -T\log P_{\rm can.}(\avphi)$, where
$P_{\rm can.}(\avphi)$ is the canonical probability distribution of
$\avphi$ for a 
particular value of $m^2_{HT}$:
\begin{equation}
  P_{\rm can.}(\avphi) \propto 
 \int {\cal D} ( A_i , \Phi ) \exp \left[- H / T \right] \,
	\delta \!\left( \avphi - \sum_x 2\Phi^\dagger\Phi/V  \right)\,.
\end{equation}
(The 2 next to $\Phi^\dagger \Phi$ is for the customary complex
normalization of the Higgs field.)
This probability distribution has to be
determined for the whole range of values from the symmetric phase to
somewhat beyond the critical bubble value $\avphiI{C}$, see
\fig\ref{maxwell3}.

In principle, $P_{\rm can.}(\avphi)$ can be calculated with a standard
lattice Monte Carlo computation, where the configurations are sampled
with the canonical probability 
\begin{equation} 
 p_{\rm can.} \propto \exp (-H/T)\,, \la{p_can}
\end{equation}
where $H$ is given in Eq.~(\ref{3D-theory}).
Algorithms for such a sampling are well known, typically taking the form
of a Markov chain in which each configuration is a relatively small
modification of the previous one.

However, in the problem we are interested in, the probability can vary
by a factor of $\sim \exp(100)$ over the range of $\avphi$ of interest.
A finite, canonical sample will simply contain {\em no} representatives
for much of the range of $\avphi$ of interest, and hence give no
information on the free energy in that part of the range; hence it will
fail to determine $P_{\rm can.}$.  For our problem, the canonical 
Monte Carlo method is utterly
useless.  This is exactly the kind of problem where multicanonical
Monte Carlo methods excel. 

In a multicanonical simulation, the Monte Carlo sampling probability
of configurations is modified so that the whole $\avphi$ range of interest is sampled 
with an approximately constant probability.
This is achieved by 
sampling the configurations according to the probability
\begin{equation}
\label{def_pmunca}
 p_{\rm muca} \propto \exp\left[ -H/T + W(\avphi)\right]\,,
\end{equation}
where the {\em weight function} $W(\avphi)$ is carefully tuned so that the
multicanonical probability distribution 
\begin{eqnarray}
P_{\rm muca}(\avphi) &\propto& 
 	\int {\cal D} ( A_i , \Phi ) \exp \left[- H / T + W \right] 
            \delta\left( \avphi - 
	\sum_x 2 \Phi^\dagger\Phi/V  \right) \nonumber \\
 	& \propto & \exp \left[ W(\avphi) \right] 
	P_{\rm can.}(\avphi) \, \la{mucaprob} 
\end{eqnarray}
is approximately constant.  This condition is met if
\begin{equation}
 W(\avphi) \approx -\log P_{\rm can.}(\avphi) + \mbox{const.}\, \la{weight}
\end{equation}
The canonical expectation value of any observable $\cal O$
can then be obtained from a multicanonically sampled set of
configurations by reweighting the individual measurements with the
weight function:
\begin{equation}
  \langle { \cal O } \rangle =
            \sum_k {\cal O}_k e^{-W(\avphiI{k})}/\sum_k e^{-W(\avphiI{k})}\,,
\end{equation}
where the sums go over all configurations in the sample.
It is not difficult to find an algorithm to perform the multicanonical
update; if one has a Markovian 
canonical Monte Carlo algorithm, application of
Metropolis accept-reject under the weight function $\exp(W(\avphi))$
after each update yields a multicanonical algorithm.  As $\avphi$ is an
easy observable to measure numerically, 
the numerical cost of this extra step is negligible. (However, since
$\avphi$ is a global quantity, some extra work is needed 
when using parallel computer architectures.)

{}From Eqs.~(\ref{mucaprob}) and (\ref{weight}) we see the main difficulty 
of the multicanonical method; we have to
know the result we are after, $P_{\rm can.}$, to some accuracy, before
we can even start the multicanonical simulation!  This requires some
kind of bootstrap process, to be discussed below, in order to
determine an initial guess for $P_{\rm can.}$; after the
multicanonical simulation, we get an improved estimate for $P_{\rm
can.}$ from Eq.~(\ref{mucaprob}).

In the above discussion we implicitly assumed that the weight
function $W$ has been optimized for one particular value of
$m_{HT}^2$.  However, due to the factorization property of the $m_{HT}^2$
term in the Hamiltonian, Eq.~(\ref{mH_pulls_out}), we obtain the
canonical probability $P_{\rm can.}$ for a whole range of
$m_{HT}^2$ values from a single multicanonical Monte Carlo run: for
example, if the multicanonical weight function has been originally
calculated with $m_{HT}^2 = m_1^2$ (and we have the resulting
distribution $P_{\rm muca}$), we have
\begin{equation}
  P_{\rm can.}(m_2^2; \avphi) \propto P_{\rm muca}(\avphi)\,\exp
  \big[-W(\avphi) + \frac{V}{2T}(m_1^2-m_2^2)\avphi\big]\,. \la{muca_reweight}
\end{equation}
Naturally we can only determine $P_{\rm can.}(m^2;\avphi)$ for values of
$\avphi$ where our $P_{\rm muca}$ determination is accurate.  
Strictly speaking, when we perform a 
multicanonical run it does not correspond to {\em any\,} particular value
of $m^2$, since we can always absorb the $m^2$ term in the action into
the weight function.

How does the computational cost scale in a multicanonical simulation?
Ideally, if we have guessed $W(\avphi)$ correctly, the
system performs a random walk in the $\avphi$ range of interest, say,
from $\avphi_1$ to $\avphi_2$.  Let us consider what happens if keep
the range of $\avphi$ considered 
fixed as we increase the volume of the system.  Now, if we
require that the system ``random walks'' through the range a
comparable number of times as the volume is increased, the
computational cost is proportional to $(\avphi_2 - \avphi_1)^2 V^2$.
The factor $V^2$ appears because $\avphi$ is an intensive variable.
For comparison, in a canonical simulation at a first order phase
transition, the numerical cost rises as ${\rm max}_{\avphi}(P_{\rm
can.}(\avphi)) / {\rm min}_{\avphi}(P_{\rm can.}(\avphi))$.  In a large
box, as previously discussed, this scales as the exponential of an
interface surface area, $\sim \exp(2\sigma V^{2/3})$, which is a
vastly more severe increase.

In realistic situations there are often ``hidden'' barriers, which can
hinder the random walk through the range of interest.  For instance, 
in our case such a 
metastability occurs in large cubic volumes 
when the surface geometry changes: as
shown in Fig.~\ref{simple}, there can be $\mbox{sphere} \leftrightarrow
\mbox{cylinder}$ and $\mbox{cylinder} \leftrightarrow \mbox{slab}$
transitions.  At the transition point two different geometries have
equal volume fractions and surface area.  During a simulation the
transition must occur by the Monte Carlo algorithm finding a series of
mixed phase geometries which smoothly 
interpolate between cylinder and sphere,
which means that the surface area must increase for a fixed volume
fraction.  Thus, if we use extremely large volumes, the transitions
between geometries become exponentially suppressed, even if the total
multicanonical probability remains constant.  (Of course for the study
of bubble nucleation we will not have to study the range of $\avphi$
where such transitions occur; but when we determine $\Teq$, or measure
the surface tension (see below), it can become an issue.  We also
observe some metastability at the value of $\avphi$ where the dominant
configuration changes from being homogeneous symmetric phase to a small
broken phase bubble.)

The success of the multicanonical method hinges on the accurate
determination of the weight function.  If we require that the
resulting probability distribution $P_{\rm muca}$ is constant up to
factor of 2, say, then the weight function must be determined up to an
accuracy of $\log(2)\approx 0.7$.  Worse accuracy in determining $W$
significantly degrades the efficiency of the subsequent Monte Carlo, so
such a requirement on the accuracy of $W$ is actually necessary.
The variation of $W$ across the
range of interest in this work is of order $\approx 100$ (that is, we
have to boost the probability of suppressed phase space regions by a
factor of $\exp(100)$.)  Thus, the weight function has to be determined
to an overall accuracy better than 1\%.

We use a continuous, piecewise linear
{\it Ansatz} for the weight function.  
We determine the weight function with an automatic iterative
calculation procedure, using variations of the procedures presented in
\cite{LaineRummukainen,broken_nonpert}.  One approach is to choose a
starting guess for $W(\avphi)$ (for instance, a constant), and to
perform a Monte Carlo under the (Markov chain) algorithm which would
generate the distribution, Eq.~(\ref{def_pmunca}).  However, after each
update sweep, $W$ is decremented at the current value of $\avphi$.
Thus, if some region of $\avphi$ is getting sampled very often, $W$ is
reduced there so it will be sampled less often.  This procedure will
cause $W$ to evolve towards the correct form, but imperfectly because
the recent history of the Monte Carlo is over-reflected in the resulting 
$W$.  To fix this, the size of the decrements is 
reduced every time the evolution successfully explores the full range of
$\avphi$ from bottom to top and back.  When the total change to $W$, in
the time the Monte Carlo evolution spans the range of $\avphi$, is
negligible, the weight function has been determined with sufficient
accuracy.  Measurements of $P_{\rm can.}$ then
consist of two parts: first, we perform a run during which the
weight function is iteratively improved 
to a required accuracy.  Second, using
this weight function, we perform a normal multicanonical run, which
gives us the final probability distribution.  The determination of $W$
typically accounts for 30--50\% of the total computational effort.

\subsection{An application: surface tension}

We illustrate how the multicanonical method works by measuring the
surface tension $\sigma$ -- i.e. the free energy/area carried by the phase
interface -- with the histogram method
\cite{Binder}.  This has become the standard and well-understood
method for computing the surface tension in a variety of lattice
theories, including work closely related to ours, SU(2) gauge + Higgs
theories \cite{KLRS_results,leipzig,desy} and effective theories for
the MSSM \cite{LaineRummukainen}.  

As discussed in Sect.~\ref{why_rare}, at the phase transition value of
$m_{HT}^2$, where the symmetric and broken phases are equally
probable, the mixed phase configurations 
with approximately equal volume fractions of
symmetric and broken phases are exponentially suppressed
(Fig.~\ref{maxwell2}).  The suppression is proportional to
$\exp(-F_{\rm surface}/T) = \exp(-\sigma \times {\rm Area}/T)$.  
This is seen as a
valley in the probability distribution of $\avphi$, see
Fig.~\ref{fig:cyldist}.

For the interface tension measurements it is advantageous to use
lattices with cylindrical geometry, $L_z \gg L_x = L_y$.  Because we
use periodic boundary conditions, there will be at least two
interfaces which span the lattice.  The cylindrical geometry makes the
interfaces tend to form parallel to the $(x,y)$-plane; and $L_z$
should then be long enough so that the two interfaces do not interact
appreciably.  This is seen as a flat minimum in the probability
distributions.

\begin{figure}[tbh]
\centerline{\epsfxsize=9cm\epsfbox{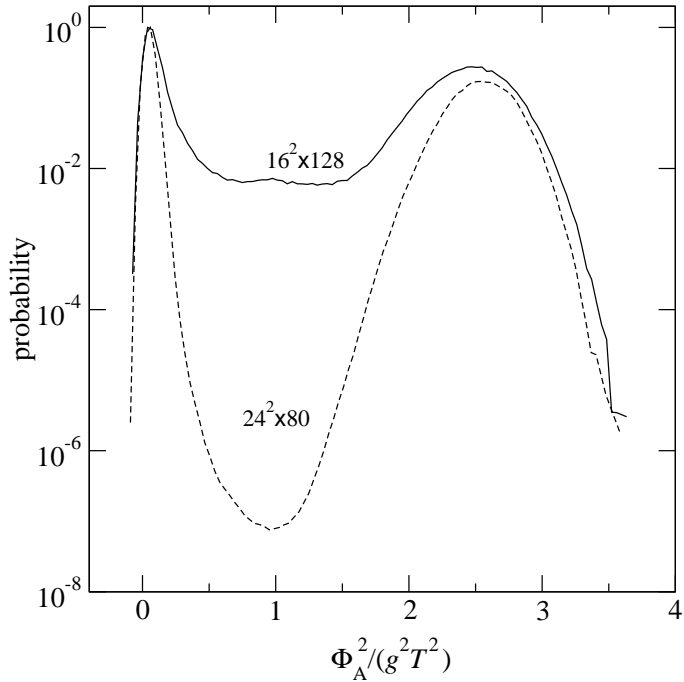}}
\vspace{0.2in}
\caption[0]{Probability distributions of $\avphi$ measured from cylindrical
lattices at $\beta = 4/(g^2 T a) = 9$ and $\lambda/g^2 = 0.036$.
\la{fig:cyldist} }
\end{figure}

We perform a multicanonical Monte Carlo with the same lattice action
and update as \cite{KLRS_results}, using the improved relation between
lattice and continuum parameters found in \cite{Oapaper}.
In Fig.~\ref{fig:cyldist} we show the probability distributions from
$16^2\times128$ and $24^2\times 80$ lattices at $4/(g^2 T a) = 9$ and
$\lambda/g^2 = 0.036$,\footnote{Corresponding to ``unimproved'' lattice
$\beta_G=9.6674$, $x\equiv \lambda/g^2=0.0389$.  For the lattice to
continuum relations for $m^2$ and $\phi^2$ see
\pcite{Oapaper}.} measured at the critical $m^2$, which is
determined by requiring that the symmetric and broken phases have
equal probabilistic weight.  (Hence, the technique also provides an
accurate determination of the the critical value of $m^2$, which we will
need for comparison later on.)  The larger lattice is not quite long
enough to have the flat central part the smaller lattice has.
Note the striking difference in widths between the symmetric (left) and
broken (right) phase peaks; this reflects the large ratio in the
$\avphi$ variance in the two phases, discussed in subsection
\ref{variance}.

The surface tension is obtained from $\log[ P_{\rm max}/P_{\rm min}
]/(2 L_x^2)\rightarrow \sigma/T$, as $V\rightarrow\infty$.  In
practice, the infinite volume value of $\sigma$ is reached in such
large volumes that finite volume analysis becomes necessary.
Following Ref.~\cite{Bunk,Iwasaki}, we fit the data with the Ansatz
which takes into account the translation modes of the surfaces and
capillary fluctuations:
\begin{equation}
  \frac{\sigma}{T} a^2 = \frac{1}{2 (L_x a)^2} \bigg[ 
            \log \frac{P_{\rm max}}{P_{\rm min}} +
            \frac{3}{2} \log L_z a - \log L_x a + \mbox{const.}\big]\,.
\end{equation}
The result of the fit from these two lattices is
\begin{equation}
  \sigma = (0.079 \pm 0.004 ) g^4 T^3 \, .
\end{equation}
We also obtain the equilibrium $m^2$, and the difference in $\avphi$
between the two phases, which in physical units is $\Delta \avphi \equiv 
\avphi({\rm broken}) - \avphi({\rm symm}) = 2.53 g^2 T^2$.

\begin{figure}[t]
\centerline{\epsfxsize=4in\epsfbox{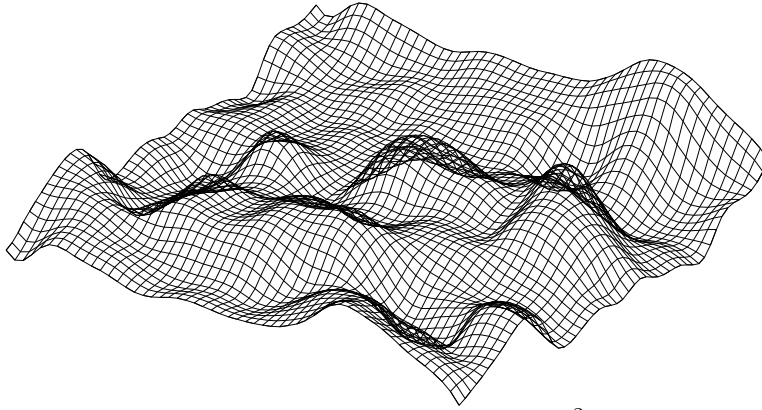}}
\caption{Geometric shape of a phase interface, from a
$62/g^2 T$ across box.  The height of the fluctuations goes as
$(T/\sigma)^{1/2}$; 
Fourier analyzing and averaging over hundreds of such
surfaces can yield an accurate determination of $\sigma$.\la{surf_pic} }
\end{figure}

\begin{table}
{\centerline{\begin{tabular}{|c|c|c|}\hline
$\quad$ Spacing $a/g^2T\quad$ & Volumes used & $\sigma$ \\ \hline
4/7  &  $72^2 \times 96$, $80^2 \times 120$, $108^2 \times 144$
& $\quad (0.0749 \pm 0.0008) g^4 T^3 \quad$ \\ \hline
4/9  &  $\quad 60^2 \times 144$, $96^2 \times 144$, $108^2 \times 160 \quad$
& $(0.0758 \pm 0.0014) g^4 T^3$ \\ \hline
4/12 &  $120^2 \times 160$, $132^2 \times 180$
& $(0.0733 \pm 0.0025) g^4 T^3$ \\ \hline
\end{tabular}}}
\vspace{0.5cm}
\caption{\la{sigma_table}Surface tension as a function of lattice
spacing.} 
\end{table}

The surface tension can be obtained much more economically
with an alternative method due to Moore and Turok \cite{MooreTurok}.
This method is based on analyzing the spectrum of the transverse
fluctuations of the phase interfaces; the magnitude of the fluctuations
is inversely proportional to $\sqrt{\sigma/T}$.  We refer the reader to
the reference for a complete discussion.
We apply this method using multicanonical tools to sample
configurations in a very large box, but now choosing $W(\avphi)$ to very
strongly prefer $\avphi$ within $5\%$ of the average between symmetric
and broken values; hence the volume always contains large regions of
each phase, with two approximately planar interfaces separating them.  We
show an example of such an interface in Fig.~\ref{surf_pic}, and present
the determined surface tension, as a function of lattice spacing, in
Table \ref{sigma_table}.  The results agree within error with the
histogram method.  If we extrapolate the values given in
the table to $a\rightarrow 0$ 
assuming $O(a^2)$ errors, as should be the case since we use an $O(a)$
corrected lattice-continuum match, we obtain the result
\begin{equation}
  \sigma = (0.0749 \pm 0.0027)  g^4 T^3 \, ,  \hspace{2cm} a = 0\, ,
\end{equation}
with a lattice spacing dependence which is small and consistent with
zero.  This indicates that our lattice spacing errors are under
control. 

\subsection{Results: probability distribution}

Let us now turn to our main problem, the determination of the
probability distribution in the region relevant to bubble nucleation,
see Fig.~\ref{maxwell3}.  The procedure is very similar to the surface
tension calculation with the histogram method described above.  
However, there are two crucial
points where it differs: 1) we need very large, preferably cubical
volumes in order for the bubbles to fit in the lattice comfortably.
The size makes it next to impossible to compute the full probability
distribution from the symmetric to the broken phase with our
computational resources.  However, 2) we need the weight function only
in the range $\avphi({\rm symm}) \le
\avphi < \left( 0.85\avphi({\rm symm}) + 0.15 \avphi({\rm broken})
\right)$, as
discussed in Sect.~\ref{finiteV}.  This guarantees that we do not yet
enter the ``cylinder'' and ``slab'' regions of the phase space, see
Fig.~\ref{simple}.  Using the random walk argument, calculating the
distribution in this restricted range only requires a factor of
$(0.15)^2 \approx 0.02$ of the resources needed for the full weight
function.\footnote{This discounts the ``barriers'' at the
bubble $\leftrightarrow$ cylinder $\leftrightarrow$ slab 
transitions, which would make the full computation even more costly.}

\begin{figure}[p]
\centerline{\epsfxsize=8.5cm\epsfbox{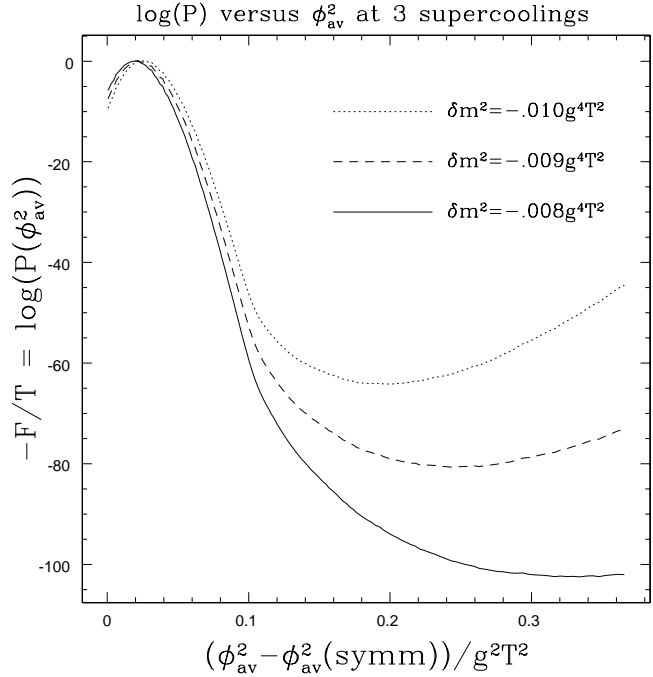}}
\vspace{0.2in}
\caption{The probability distribution for $\avphi$ at three values of
$\delta m^2$, $-0.008g^4T^2$ (solid), $-0.009g^4T^2$ (dashed), and $-0.010
g^4T^2$ (dotted), for $124^3$ lattice at $g^2aT=4/9$.  
In each case the local maximum is the supercooled
symmetric phase, the local minimum is the critical bubble.  The three
curves are obtained by reweighting the same multicanonical run.  Greater 
supercooling leads to less suppression of the critical bubble.
\la{fig:bubdist}}
\end{figure}

\begin{figure}[p]
\centerline{\epsfxsize=5in\epsfbox{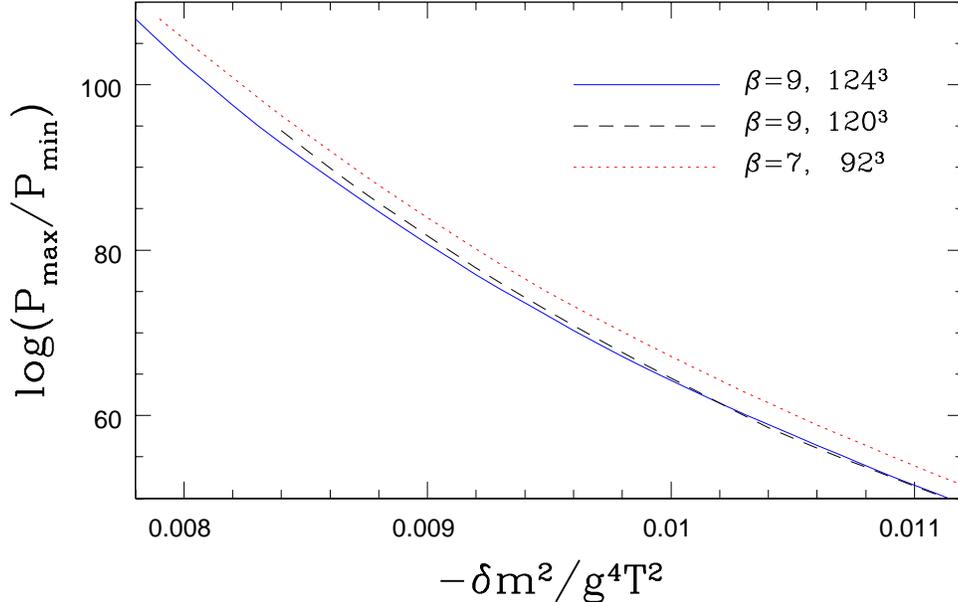}}
\caption{\la{resultfig} Log of ratio of probabilities $\log P_{\rm
can.}({\rm meta}) / P_{\rm can.}({\rm crit})$, as a function of $\delta
m^2$ the supercooling from the equilibrium $m^2$.  The solid
%, blue 
line
is the $124^3$ lattice, and the dashed
%, black 
line is the $120^3$
lattice, at $(4/g^2Ta)=9$.  Each has a statistical error of $\pm 1$, so
they agree within expected error.  The 
%red, 
dotted line is the data from
the $92^3$ lattice at $(4/g^2Ta)=7$.  Its disagreement represents
definite, but small, lattice spacing error.}
\end{figure}

In Fig.~\ref{fig:bubdist} we show the probability distributions from a
$124^3$ lattice, with $4/(g^2Ta) = 9$.  The result of a single multicanonical
run has been reweighted to 3 different values of $m_{HT}^2$, namely
$(0.008,0.009,0.010)g^4 T^2$ below the equilibrium value.  The critical
bubbles correspond to the minimum locations of the probability.  Note
that the value of $\avphi$ for the critical bubble moves to smaller
values as we increase the supercooling; this is because the critical
bubble gets smaller at larger supercooling, and $\avphi$ is a volume
average.  We
also have results from a $92^3$ box at $4/(g^2Ta)=7$, to study lattice
spacing dependence, and from a $120^3$ box at $4/(g^2Ta) = 9$, which
gives weak information on volume dependence.  The ratio of
probabilities between the metastable minimum and the critical bubble,
$P_{\rm can.}(\avphi = \avphiC) / P_{\rm can.}(\avphi=\avphi({\rm symm}))$,
is plotted for each lattice as a function of $\delta m^2 \equiv
m_{HT}^2 - m_{HT}^2({\rm equilib})$ in
Fig.~\ref{resultfig}.  This figure corresponds to the difference in
height, in Fig.~\ref{fig:bubdist}, between the local maximum and local
minimum.  For each curve in the figure, the statistical error bars are
about $\pm 1$, with strong correlation in the error along the curve.
The finer spacing lattices agree within errors.
This is a (weak) check on volume dependence, but it is also a check 
of the code, since the $120^3$ and $124^3$ volume
computations were performed with completely independent sets of code, on
machines of different architecture.  
The coarser lattice data differs by between $2\%$ and $3\%$.  This probably
represents lattice spacing errors; if we extrapolate assuming $a^2$
errors (the first order which should be present, due to our lattice
improvement), we estimate the difference between the finer lattice and
the continuum is $1.5$ times as large as the difference in the curves.
Where the ratio of probabilities is $e^{100}$, this would be a
correction of $e^4$ in the nucleation rate, 
or about a $2.5\%$ correction in $\delta m^2$.
It remains to integrate the area in the
symmetric phase minimum, to compute the dynamical contributions, and
divide by the volume, to convert this result into the real time rate.

\subsection{Dynamical prefactor, tools and calculation} \la{sec_dyn}

To determine the real time rate for nucleations, we now have to perform
real time evolution on each of a sample of configurations with $\avphi = 
\avphiC$.  We get the sample of such configurations by, first, choosing
a $\delta m^2$ to consider, and, next, by performing a
multicanonical Monte Carlo, as just described, and recording those
configurations for which $\avphi$ lies within a narrow tolerance of
$\avphiC$ (which depends on $\delta m^2$, see Fig.~\ref{fig:bubdist}).  
In fact we can speed up the sampling process by choosing a
weight function $W(\avphi)$ which favors $\avphi = \avphiC$ even more
strongly than the one used to determine the probability distribution in
the last section.  Then, we must study the real time evolution of each
configuration in the sample, at the thermal Higgs 
mass $m^2_{\rm eq} - \delta m^2$. 

As discussed in subsec.~\ref{dyn_subsec}, the appropriate dynamics, at
leading log, are Langevin dynamics.  The gauge fields evolve according
to Eq.~(\ref{Bod_th}), and the Higgs fields also evolve under Langevin
dynamics, with a much (parametrically) faster time scale.  In continuum
notation, this means evolving the fields under the following Langevin
field equations (normalizing $E$ so
$E^2/2g^2$ appears in the action),
\begin{eqnarray}
\label{really_use}
\sigmaE E_i^a(x) & = & - \frac{\partial}{\partial A_i^a(x)}
	H + \xi_i^a(x,t) \, , \\
\sigmaE D_t \Phi(x) & = & - \eta \frac{\partial}{\partial
	\Phi^\dagger(x)} H + \xi_\Phi(x,t) \, , \nonumber \\
\langle \xi_i^a(x,t) \xi_j^b(x',t') \rangle
	& = & 2 \sigmaE T \delta_{ij} \delta^{ab}
	\delta(x-x')\delta(t-t') \, , \nonumber \\
\langle \xi_\Phi(x,t) \xi^\dagger_\Phi(x',t') \rangle 
	& = & 2 \eta \sigmaE T 
	\; {\bf 1} \delta(x-x') \delta(t-t') \, , \nonumber
\end{eqnarray}
where $H$ is given in Eq.~(\ref{3D-theory}),
${\bf 1}$ is the identity in component space for the Higgs field
and $\eta$ is the ratio of the speeds of Langevin evolution, which
is parametrically $\sim 1/\alpha_w$, and so 
should be taken large.  

It is possible to perform numerical Langevin evolution on
lattice fields, but it is slow and unnecessary; any dissipative update
will do, if the relation between the number of updates and Langevin time 
is known.  Hence we use the heat bath algorithm to update the gauge
fields; the relation between the number of heat bath updates, 
and the time scale $t$ in the equations above, is
discussed at some length in \cite{Bodek_paper}.  At leading order in
small $a$ the
relation is that, for random order heat bath updates of the lattice
links, $n$ updates per link corresponds to $\Delta t = a^2 \sigmaE n/4$.
(Note that this relation is specifically for our choice of lattice
action, namely ``Wilson glue''; it would be different for an improved
action.  It can also differs if the sites are updated in a specific, rather 
than random, order, and in fact depends on the order of update.)  
We update the Higgs fields with
a mixture of the over-relaxation algorithm presented in
\cite{KLRS_results} and a Higgs field heat bath algorithm.  Note that
our ``real time'' evolution algorithm can be viewed as a canonical
Monte Carlo evolution algorithm at $m^2 = m^2_{\rm eq}-\delta m^2$;
hence there is no concern that it somehow spoils the thermodynamics (as
might happen for Langevin or Hamiltonian evolution with a finite time
step, due to time step size errors).

As described in subsection \ref{real_time}, we calculate
the ``dynamical prefactor'' $\d$, Eq.~(\ref{ddef}), by evolving
a critical bubble configuration both forward and backwards in time,
long enough to see whether the system evolves either towards the
symmetric or the broken phase.  Since forward and backwards Langevin
evolution are equivalent, in practice we generate a few Langevin
trajectories from each initial critical bubble configuration; each pair
can be joined 
together to form a full trajectory, see Fig.~\ref{trajectory}, and by
considering all pairings we somewhat improve the statistics.
The dynamical prefactor is then the expectation value
\begin{equation}
 \d = \frac{1}{N_{\rm traj.}} \sum_{\rm traj.} \,
	\frac {\delta_{\rm tunnel} }{\mbox{\# of crossings} }\,,
\end{equation}
where $\delta_{\rm tunnel}$ is 1 if the trajectory leads to tunnelling,
0 otherwise, and (\# of crossings) is the number of times the
trajectory crosses the critical bubble value of $\avphi$.  When
computing the error in the determination of $\d$ we must account
for the dependence of the several trajectories involving a common
critical bubble.

\begin{figure}[tb]
\centerline{\epsfxsize=5in\epsfbox{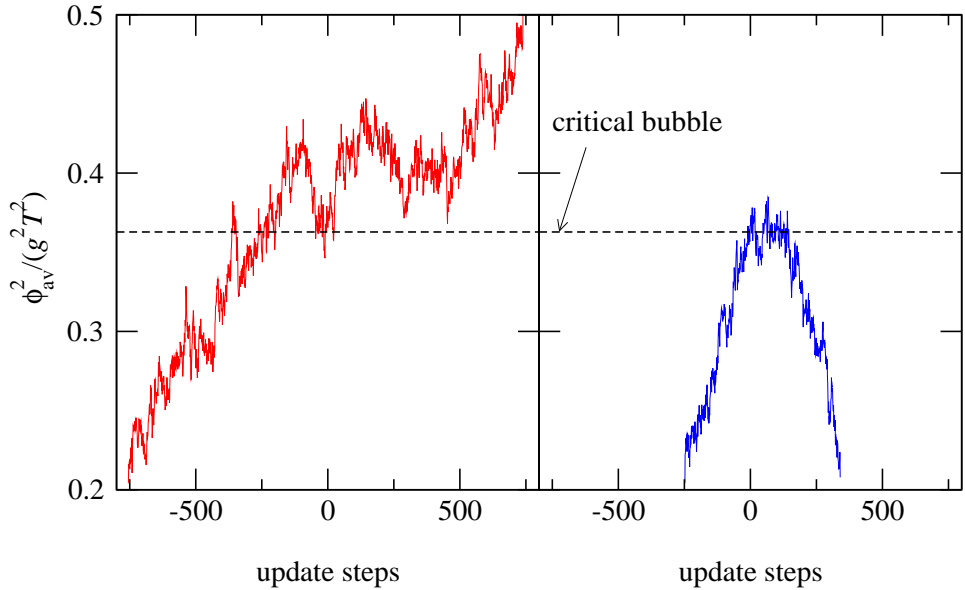}}
\vspace{8mm}
\caption{\la{trajectory}A tunnelling trajectory and a trajectory which 
does not lead to tunnelling, measured from $120^3$ volume and at
supercooling $\delta m^2 = -0.0082\,g^4 T^2$.  The horizontal dashed
line is the critical bubble value of $\avphi$.  Two half-trajectories
(positive and negative timestep values) are evaluated starting from a
configuration at the critical value of $\avphi$, and glued together at
timestep 0 to form a full trajectory.  }
\end{figure}

One legitimate concern is that the Langevin dynamics we consider 
are only correct at
leading log, and as argued in \cite{Bodek_higgs}, the treatment should
break down when there is a large Higgs field condensate.  Hence our
treatment of the dynamics may not be better than an $O(1)$
treatment.  However, our determination of the thermodynamic likelihood
of a critical bubble was only good to $\pm 1$ in the exponent, so an
$O(1)$ error in the dynamics is no worse; in any case, since the
nucleation rate is $\sim e^{-100} T^4$, a factor of 2 error in its
determination only represents a $1\%$ error in the exponent, and
somewhat less than a $1\%$ error in the determination of $\delta m^2$,
as seen in Fig.~\ref{resultfig}.  It would be possible to do a better
job by using the full HTL dynamics by the technique developed in
\cite{BMR}; however, this approach is much more numerically expensive.
It also requires the inclusion of the $A_0$ field in the thermodynamic
treatment, and it could be difficult to eliminate finite time step
errors in the dynamics, which could make the thermodynamics explored by
the evolution slightly incorrect.

\begin{figure}[p]
\centerline{\epsfxsize=5in\epsfbox{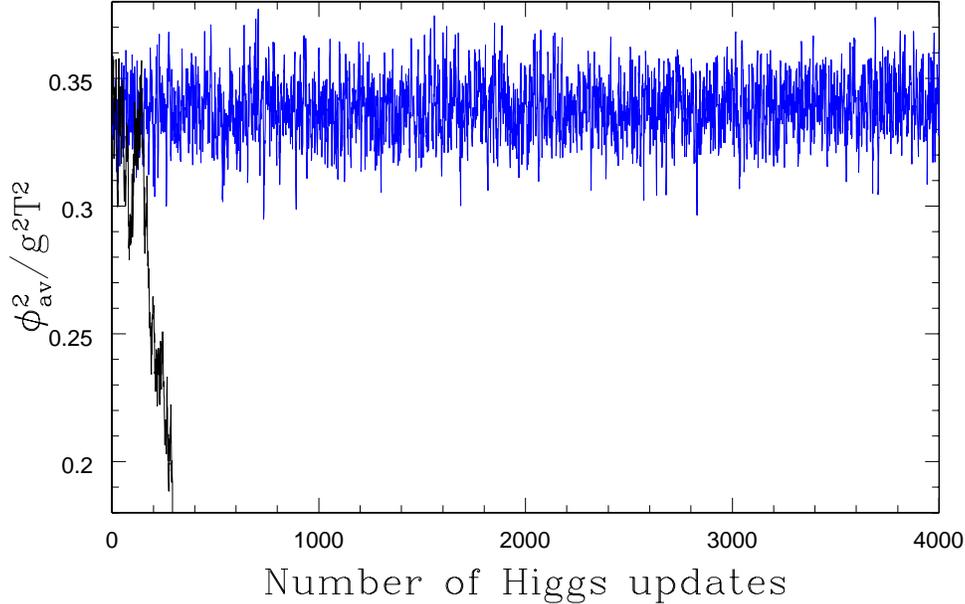}}
\vspace{0.15in}
\caption{\la{frozen} Two time histories in the $124^3$ volume, 
both starting at the critical
bubble:  that in which the gauge field is not updated (blue, stays
nearly constant) and that in which the gauge field is updated (black).
When both fields are updated, the critical bubble can grow, or, in this
case, collapse; if only the Higgs fields are updated it does not
evolve.} 
\end{figure}

The reader might also be concerned that the dynamics of a broken phase bubble 
will not have a good limit as $\eta$ is taken large.  If the Higgs
fields are allowed to evolve much faster than the gauge fields, won't
the bubble either collapse or expand on a time scale set by the rate of the 
Higgs field evolution?  The answer is, no.  If we choose a starting
configuration with a broken phase bubble, and we evolve the Higgs fields 
without allowing the gauge fields to evolve, the bubble does {\em not}
collapse but remains a critical bubble indefinitely.  It is essential
that the electroweak phase transition is a radiatively induced phase
transition; the state of the gauge field fluctuations, alone, are
sufficient to indicate which phase a configuration 
is in, and the Higgs condensate
cannot expand into the symmetric phase, or collapse inside the bubble, 
without the gauge field fluctuations changing as well.  To demonstrate
this point, we have performed an $\eta = \infty$ evolution, meaning an
evolution in which only the Higgs fields, and not the gauge fields, are
allowed to evolve.  It is compared to an evolution in which both evolve, 
but the gauge fields evolve much more slowly, in Fig.~\ref{frozen}.
Naturally, the value of $\avphi$ changes during each evolution; but in
the evolution with frozen gauge fields, it just bounces around a central 
value, and the bubble is stable.  We should expect, then, that
$\avchange$ and $\d$ will depend on $\eta$.  But, as we must check,
the product should have a finite limit.  We would also like to know how
the product $\avchange \: \d$ depends on $\delta m^2$; we expect that
the dependence is weak.

To check this we have used two values of $\eta$, $\eta=5$ and $\eta=10$, 
and measured $\d$ and $\avchange$ for each (on a $124^3$ lattice at 
$\delta m^2=-0.0082g^4 T^2$).  The results are
presented in Table \ref{dyn_table}, which shows that the product ${\bf
d} \avchange$ is independent of $\eta$, within
numerical errors.  We have also re-analyzed the same set of trajectories,
sampling $\avphi$ half as often;
we find as expected that $\d$ and $\avchange$ each depend
strongly on the sampling rate, but the product does not.  

\begin{table}[p]
\centerline{\begin{tabular}{|c|c|c|c|c|} \hline
$\quad \eta \quad$ & 
$\displaystyle \; {\vphantom{\Bigg( )}}\frac{4\, \Delta t}{\sigmaE a^2}\;$ & 
$\; \displaystyle \dispavchg \; \frac{\sigmaE}{\alpha_w^3 T^4}\;$ 
& $\d$ & 
$\; \displaystyle \dispavchg\, \d\; \frac{\sigmaE}{\alpha_w^3 T^4}\;$ 
\\ \hline
5 & 1  & $366\pm\; 8 $&$\;0.0169 \pm 0.0028\;$ & $6.2 \pm 1.0$ \\ \hline
5 & 2  & $240\pm\; 7 $& $0.0264 \pm 0.0048$  & $6.3 \pm 1.2$ \\ \hline
10 & 1 & $500\pm 10$  & $0.0146 \pm 0.0037$  & $7.3 \pm 1.8$ \\ \hline 
10 & 2 & $322\pm\; 6$ & $0.0218 \pm 0.0048$  & $7.0 \pm 1.6$ \\ \hline 
\end{tabular}}
\vspace{0.08in}
\caption{\la{dyn_table}
Dynamical information $\avchange$ and $\d$, varying
$\eta$ and sampling each data series at two rates (all on a $124^3$
lattice at $\delta m^2=-0.0082g^4T^2$).  Each pair of data at 
fixed $\eta$ comes from the same set of trajectories; 
comparison shows that ${\bf
d}$ and $\avchange$ depend strongly on sampling frequency, but the
product does not.  The (fully independent) data sets with different
$\eta$ agree within error for $\avchange \, \d$, showing this
quantity has good large $\eta$ behavior.}
\end{table}

\begin{table}[p]
\centerline{\begin{tabular}{|c|c|c|c|c|c|} \hline
$\displaystyle \frac{\delta m^2}{g^4T^2} $ & 
$ -\log [P_{C}\times \alpha_w T^2] $ 
& $\displaystyle \dispavchg 
    \frac{\sigmaE}{\alpha_w^3 T^4}$ 
& $ \vphantom{\Bigg|} \d$ 
& $-\log\bigg[\frac{\mbox{rate}}{V} 
	\frac{\sigmaE / T}{\alpha_w^5 T^4}\bigg] $ \\ \hline
-0.00835 &$ 94.1\pm0.6 $&
$343\pm 6 $&$0.025 \pm 0.008$&$ 97.0 \pm 0.7 $ \\ \hline
-0.00951 &$ 71.0\pm0.5 $&
$331\pm 9 $&$0.020 \pm 0.005$&$ 74.1 \pm 0.6 $ \\ \hline
-0.01057 &$ 54.8\pm0.6 $&
$314\pm 8 $&$0.009 \pm 0.004$&$ 58.8 \pm 0.7 $ \\ \hline 
\end{tabular}}
\vspace{0.08in}
\caption{\la{dyn_table2}%
The nucleation rate calculated with 3 different supercoolings, in a
$120^3$, $g^2 T a = 4/9$ lattice.  $P_C$ is the probability density that 
$\avphi = \avphiC$, calculated from Eq.~(\ref{flux_2}).  The
nucleation rate in the fifth column is $P_{C}\avchange \d /2$.}
\end{table}

We have also varied the degree of supercooling, to confirm that the
dependence of the dynamical prefactors is not very strong.  In the
$120^3$ box we have results at 3 different values of supercooling:
$\delta m^2/(g^4T^2) = -0.0082$, $-0.0093$ and $-0.0104$.  These
correspond to significantly different bubble probabilities, as shown
in Table \ref{dyn_table2}: the largest supercooling corresponds to
bubbles which are $\approx e^{39}$ times more likely to nucleate than
with the smallest supercooling.  The bubble probability density,
Eq.~(\ref{flux_2}), is readily evaluated by integrating the data shown
in Fig.~\ref{fig:bubdist}.  However, as expected, the dynamical
factors are seen to be fairly stable throughout this region, varying
only by approximately a factor of two or three, which is in practice
insignificant when compared with the differences in probability.

Note the units on $\avchange$.  The 
units on the first term in Eq.~(\ref{flux_eq}) are the same as $1/\avphi 
\sim 1/\alpha_w T^2$; while to get a rate per unit volume we must divide 
by the volume, $1/V \sim \alpha_w^3 T^3$.  Hence the parametric
appearance of the nucleation rate is $\propto \alpha_w^5 T^4 \log(1/g)$, 
using $\sigmaE \sim T/\log(1/g)$.  This arises simply from the relation
between dimensionless lattice quantities and physical quantities.

\subsection{Numerical results and relation to cosmology}

The previous two subsections contain all the ingredients needed to
determine the real time rate for bubble nucleation.  It remains, first,
to put the ingredients together, and second, to determine what value for 
the nucleation rate is interesting cosmologically. 
We write the bubble nucleation rate as
\begin{equation}
\frac{{\rm rate}}{V} = \left( \frac{g^2 T^2}{m_D^2} \right)
	\log(1/g) \alpha_w^5 T^4 \exp(-S) \, .
\end{equation}
For the realistic standard model values $m_D^2 = (11/6)g^2 T^2$ and
$\alpha_w \simeq 1/30$, minus the log of the term in front (evaluating
$\log(1/g)$ using Eq.~(\ref{Bod_th})) is 16, 
so the rate is $\exp(-(S+16))T^4$.
Our result for $S$ is shown, as a function of $\delta m^2$, in Figure
\ref{final_fig}.  This represents our final numerical result.  It is
unfortunate that the numerical effort is too large to perform the
calculation for several couplings, and we have also not considered a
realistic set of parameters in the MSSM.

\begin{figure}[t]
\centerline{\epsfxsize=5.5in\epsfbox{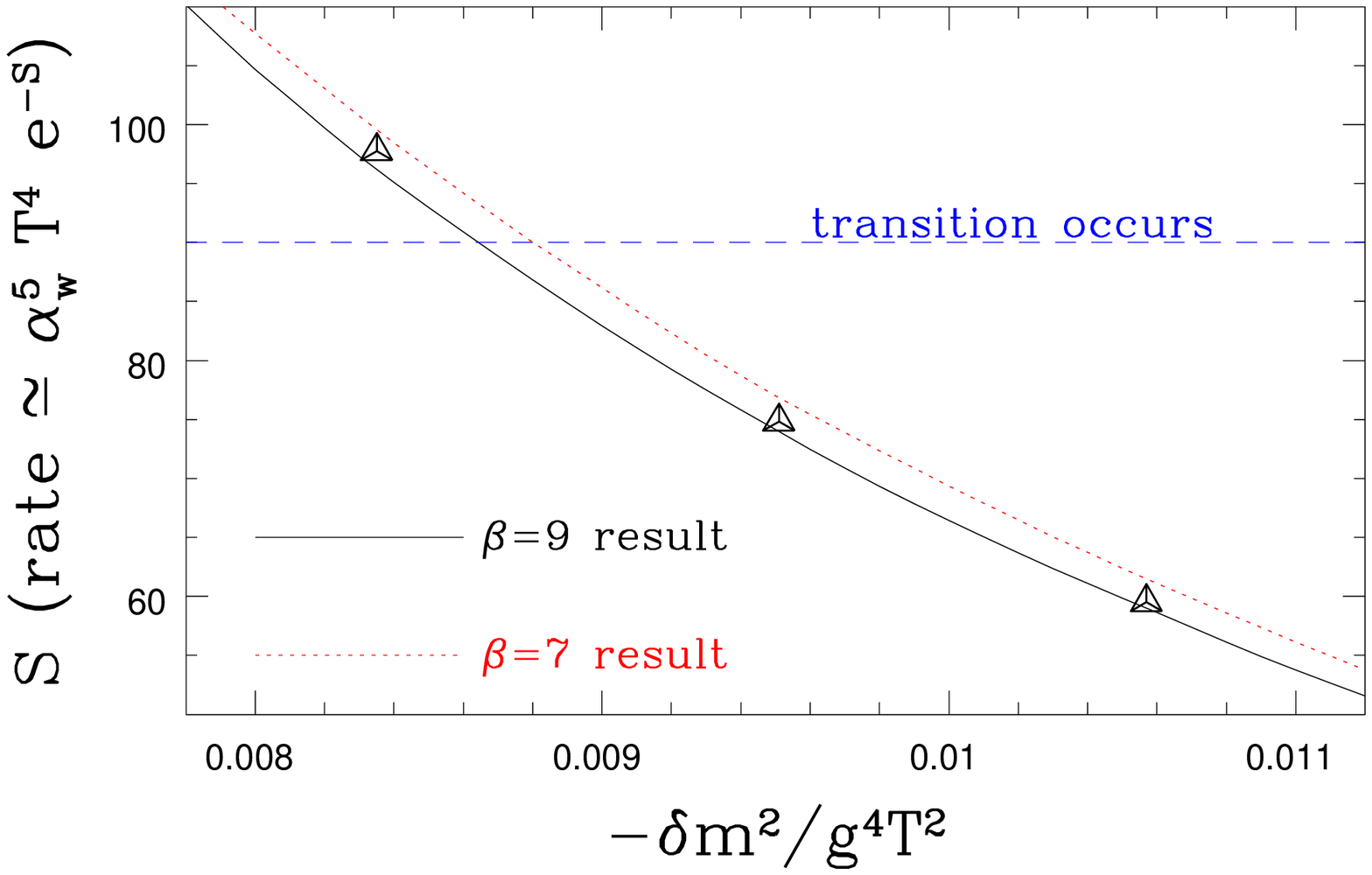}}
\vspace{0.1in}
\caption{\la{final_fig}
$S$, as defined in the text, as a function
of $\delta m^2$, the degree of supercooling, for the $g^2 T a=4/7$
(dotted) and $g^2 T a=4/9$, $124^3$ lattice (solid) data, each using
$\avchange \d$ as evaluated at $\delta m^2=-0.0082 g^4 T^2$.  The
points are the $120^3$ data, using the three evaluations of $\avchange
\d$ at the values of $\delta m^2$ where they were evaluated.  
Error bars, not shown, are 
dominated by the error in the determined probability
distribution, and are about $\pm 1$ in $S$.}
\end{figure}

What are the errors of $S$ in Figure~\ref{final_fig}?  Using the
numbers in Table~\ref{dyn_table2}, for example, we see that the final
statistical errors are strongly dominated by the errors of the
probability distribution $P$.  In addition, there is the systematical
error of the real-time update evolution (see sect. \ref{sec_dyn}).  Both
of these error sources are easily included by a $\pm 2$ error band around $S$
in Figure~\ref{final_fig}.

We want to know at what value of $S$ the phase transition occurs, so we
can determine $\delta m^2$ and therefore the amount of supercooling.
The relevant picture is discussed in \cite{EIKR}.  The bubbles of broken 
phase which convert most of the volume into the broken phase, nucleate
over a characteristic period of time $t_{\rm nuc}$, and with a mean
separation $d_{\rm nuc}$, which is also the diameter they grow to.  For
one such bubble to nucleate per $d_{\rm nuc}^3$ volume in time $t_{\rm
nuc}$, the nucleation rate must be $\sim 1/(t_{\rm nuc} d_{\rm nuc}^3)$.
The time scale is set by how fast the nucleation rate is changing;
$t_{\rm nuc} \simeq ( dS/dt )^{-1} \simeq ( H(T) T dS/dT )^{-1}$, with
$H(T)$ the Hubble's constant at temperature $T$.  The separation is 
$d_{\rm nuc} \sim v_{\rm s} t_{\rm nuc}$, with $v_{\rm s}$ the sound
speed $\sim 1$; this is because a bubble is preceded by a shock front
propagating at approximately $v_{\rm s}$, which heats the plasma, and
nucleations are suppressed in the heated region.  Hence the nucleations
take place when
\begin{equation}
{\rm rate} \simeq \left( H \frac{T dS}{dT} \right)^{-4} \, .
\end{equation}
{}From Fig.~\ref{final_fig}, and from the mass-temperature relation,
Eq.~(\ref{m_of_T}), we see $TdS/dT \sim 2\times 10^4$; while from the
Friedman equation, $H =\sqrt{4\pi^3 g_*/45} (T^2/m_{\rm pl}) \sim
e^{-36.5}T$. Hence the interesting value of $S$ is $16+S \simeq 106$, or
$S \simeq 90$.  The $g^2Ta=4/7$ data show this value is obtained at
$\delta m^2 = (-0.00880 \pm 0.00010) g^4 T^2$, while the finer lattice
data give $(-0.00864 \pm 0.00010) g^4T^2$.  An extrapolation to zero
lattice spacing, assuming $O(a^2)$ errors, gives 
$\delta m^2= -0.00840 \pm 0.00020$.  Again, this is for $\lambda/g^2 =
0.036$; naturally the amount of supercooling is strongly dependent
on the strength of the phase transition, and hence on $\lambda / g^2$.  
This is our final numerical result.

%------------------------------------------------------------------
\section{Other Approaches}
\la{otherways}

Since we have only computed the nucleation rate nonperturbatively for
a single set of parameters, the main application of this work is as a
benchmark for studying the performance of other, less first principles
means of determining the nucleation temperature, which have previously
been used in the literature.  For this purpose, we will apply a few of
these techniques to the current set of parameters, to see how
accurately they determine $\delta m^2$.  (This actually gives an
optimistic appraisal; the dependence of $S$ on $\delta m^2$ is strong,
so the relative error in $S$ the log of the nucleation rate is
typically almost twice the relative error in $\delta m^2$.)

\subsection{Thin wall approximation estimate}

The idea of the thin wall approximation is to hope or assume that the
critical bubble is ``thin walled,'' which technically means we assume 3
things:
\begin{enumerate}
\item the bubble radius is much greater than the thickness of the phase
interface, so we can treat the interface as an infinitely thin geometric 
surface;
\item the surface tension of the interface, at the nucleation
temperature, is the same as at the equilibrium temperature;
\item the free energy difference between the phases is given by the
leading order expression $\Delta V = l \Delta T/T$, with $l$ the latent
heat at equilibrium.  This neglects the change in $l$ between the nucleation 
and equilibrium temperatures.
\end{enumerate}
In the limit of small supercooling, all three approximations become
exact.  Hence the thin wall approximation is appropriate if we are
interested in very small $\delta m^2$, where the critical bubble
free energy is huge.  The approximation is in expecting it to continue to
work down to where the critical bubble free energy is $S \sim 90 T$.

The thin wall approximation is that a bubble of radius $r$ will have
energy 
\begin{equation}
E(r) = 4 \pi r^2 \sigma - \frac{4 \pi}{3} r^3 \Delta V \, ,
\end{equation}
with $\sigma$ the equilibrium surface tension and 
\begin{equation}
\Delta V = \frac{l_{\rm eq} \Delta T}{T} = 
	\frac{ -\delta m^2}{2} \Delta \avphi(\Teq) \, .
\end{equation}
We find the extremum over all $r$, 
\begin{equation}
\frac{\partial E}{\partial r} = 0 \quad \Rightarrow \quad
r = \frac{2 \sigma}{\Delta V} \, , \quad
E = \frac{16 \pi \sigma^3}{3 (\Delta V)^2} \, .
\end{equation}
We now estimate that the nucleation rate will be
\begin{equation}
{\rm rate} \: = \exp(-E/T) \left( \frac{g^2 T^2}{m_D^2} \right)
	\alpha_w^5 \log(1/g) T^4 \, ,
\end{equation}
where the technique really provides us no way of getting the
non-exponential term we write, but we guess that this is correct on
parametric grounds.  

For comparison with the case we have studied numerically, we solve for
the value of $\delta m^2$ required to make $E/T=S=90$, using the
nonperturbatively determined values of $\sigma$ and $\Delta
\avphi(\Teq)$ from the
last section, $\sigma = 0.075 g^4 T^3$ and $\Delta \avphi(\Teq) = 2.53
g^2 T^2$. Substituting into 
\begin{equation}
-\delta m^2 = \left( \frac{64 \pi \sigma^3}
	{3(\Delta \avphi(\Teq))^2 E} \right)^{1/2}
\end{equation}
gives $\delta m^2 = -0.0070 g^4 T^2$.  This is low by about $20\%$
from the nonperturbative value.

Alternately, we could ask how accurately the thin wall approximation
predicts the nucleation rate at the actual value of $\delta m^2$.
Plugging in the value of $\delta m^2$ found in the last section gives 
$S = 62$, which is off by almost a third.

It is clear that the thin wall approximation is doomed to errors of
this magnitude when applied where $E/T \sim 90$.  Note that the radius
of the critical bubble, for $\delta m^2$ determined above, is
\begin{equation}
r = \left( \frac{3 E } { 4 \pi \sigma } \right)^{1/2} \, .
\end{equation}
But it is impossible for the thickness of the interface to be less than
$\sim \sqrt{T/\sigma}$; even if the intrinsic thickness were somehow thinner, 
the surface fluctuations would generate such a thickness.
Hence $r$ never greatly exceeds
the surface thickness if we are interested in $E/T \sim
90$.  However, all considering, the thin wall approximation (given
nonperturbative inputs) does pretty well.

We should comment that, although the thin wall approximation requires
nonperturbative inputs, it is much easier to determine $\Delta \avphi$ and
$\sigma$ on the lattice than to directly compute the nucleation rate.
In particular, we could think realistically 
of doing a ``scan'' of $\Delta \avphi$ and
$\sigma$ at several parameter values, in the standard model or one of
its extensions, but to do the nucleation rate at numerous values of
parameters would probably be beyond what is currently a reasonable
amount of numerical effort.  Hence the thin wall approximation may be a
reasonable approach to getting ``rough and ready'' nucleation
information.

\subsection{Perturbative estimate}

The traditional method for determining the bubble nucleation rate in the 
context of a perturbative treatment of the strength of the phase
transition (see for instance \cite{EIKR,Turok,Dine}) is to approximate the
free energy (effective action) to be the tree kinetic term for the
(gauge fixed) Higgs condensate $\phi$, plus an
effective potential term computed at some order in the loop expansion, 
\begin{equation}
E = \int d^3 x \left( \frac{1}{2} (\nabla \phi)^2 + V(\phi) \right) \, . 
\la{E1}
\end{equation}
It is also possible to consider radiative corrections to the Higgs field 
kinetic energy, as we discuss in the next subsection.%
\footnote{We note that a full-fledged analytical computation of the
nucleation rate in the spirit of the Langer method is very difficult
in radiatively induced first order transitions.  The problem is how to
distinguish the fluctuations which give you the effective potential
from the fluctuations of the bubble in this potential.  The results
from a cubic anisotropy model (a simple spin model) calculation by
Strumia and Tetradis \pcite{ST} display a dramatic dependence on how
the separation of the fluctuations is done; since the physics cannot
depend on the separation of fluctuations, this effect must be an
artifact of the approximations done in the calculation.  On the other
hand, if there is a sufficiently strong first order transition already
at tree level, the Langer method is relatively straightforward to
apply reliably \pcite{Munster,BTW,MST}. }

We begin with the simplest possible estimate.  We take the one loop
effective potential, Eq.~(\ref{1loop}), reproduced here for convenience:
\begin{equation}
V_{\rm 1 \; loop} = \frac{m_{HT}^2}{2} \phi^2 
	- \frac{g^3}{16 \pi} \phi^3 T + \frac{\lambda}{4} \phi^4 \, .
\la{1loop2}
\end{equation}
Here we have neglected scalar loops, which give a contribution down
relative to the $\phi^3$ term by $(\lambda / g^2)^{3/2}$.
We can see from Eq.~(\ref{1loop2}) that at 
\begin{equation}
m_{HT}^2 = m_{\Teq}^2(1 \; {\rm loop}) 
	= \frac{g^6 T^2}{128 \pi^2 \lambda}
\end{equation}
there are two degenerate minima with
\begin{equation}
\phi=0 \qquad {\rm and} \qquad 
	\phi = \phi_0(1 \; {\rm loop}) = \frac{g^3 T}{8 \pi \lambda} \, .
\end{equation}
At best perturbation theory is an expansion in $\sim ( g^2 T/m_W ) \sim
gT/\phi$, hence in $\lambda /g^2$.  Two loop corrections will give a
correction to $\phi_0$ of order $(\lambda /g^2) \phi_0$, which is larger
than the scalar loop contribution we neglected above.  This justifies
neglect of the scalar loop.\footnote{As an aside we mention that the
expressions usually 
written for scalar loops (refs. \pcite{Hebecker,FKRS1}, though
not ref. {\pcite{Hebecker2}})
must be inconsistent because they depend on $m_{HT}^2$ 
in a way which does not satisfy Eq.~\protect{\ref{mH_pulls_out}}.
The problem is that they were derived using a Higgs mass in loops which
does not correspond to the curvature
of the potential; the value of $m_H^2$ used in loops 
includes the second derivative of the $\lambda \phi^4$ potential term but
not of the negative cubic term, which is of the same size in the
interesting region and cannot be neglected in any consistent approximation 
scheme.  If we include terms in the effective potential in the spirit of 
an expansion in $\lambda / g^2$ then no reference to the scalar
self-coupling or $m_{HT}^2$ appears in any loop induced term
until $O((\lambda / g^2)^{3/2})$,
when an infinite class of diagrams must be resummed, corresponding to a
single Higgs loop but including iterated one loop
mass corrections from gauge 
boson loops within the 3D theory (not just the nonzero Matsubara
frequency loops).\la{rant}}

One then assumes, reasonably, that the critical bubble, the saddle point
of Eq.~(\ref{E1}), will have spherical symmetry, and looks for the
saddle point of
\begin{equation}
E = 4 \pi \int_0^\infty r^2 \left( \frac{1}{2} (\partial_r \phi(r))^2 
	+ V(r(\phi)) \right)
\la{E2}
\end{equation}
over all $\phi(r)$, with the boundary condition that $\phi(r)$ go to
zero at large $r$, so we have a bubble in the symmetric phase.  One
finds the saddle point free energy as a function of $m_{HT}^2$ and looks for
where it takes on the desired value, say $E/T=S=90$.  The difference from
the equilibrium $m_{HT}^2$ is the degree of supercooling we are after.  
(As in the previous section we will simply assume that the bubble zero
modes etc.~give a dimensionful prefactor of $\sim \alpha_w^5 T^4$.
Since even a
change in the numerical value of this prefactor by a factor of 1000
would represent only a change of $7$ in the log of the nucleation rate,
our results depend only weakly on this treatment.)

We are not aware of a purely analytic way to find the
saddle point of Eq.~(\ref{E2}), but the well known overshoot-undershoot
algorithm is both efficient and accurate.  We find, for $\lambda / g^2 = 
0.036$, that
\begin{equation}
\delta m^2 = -0.0058 g^4 T^2 \, , \qquad
	\sigma_{\rm equilibrium} = \int_0^{\phi_0} \sqrt{2V}d\phi
	= \frac{g^9T^3}{3072 \sqrt{2} \pi^3 \lambda^{5/2}}
	= 0.0302 g^4 T^3 \, .
\end{equation}
In comparison to the nonperturbative values these are both fairly far off.
The degree of supercooling is $2/3$ of the right value and the surface
tension is $40\%$ of the nonperturbative value.  One loop perturbation
theory is NOT accurate even at $\lambda /g^2 = 0.036$, though it is not
completely wrong.

We therefore go on to the two loop effective potential.  
Neglecting powers of $\lambda$ and setting $m_{HT} \ll m_W$ to zero 
within loops, the new
term in the effective potential is \cite{ArnoldEspinosa}
\begin{equation}
V_{2 {\rm \; loop}} = V_{1{\rm \; loop}} - \frac{51}{512 \pi^2}
	g^4 \phi^2 T^2 \log(\phi/gT) \, ,
\end{equation}
where the choice of $gT$ inside the logarithm is a convenient choice of
renormalization point; a different choice can be absorbed into a shift
in $m_{HT}^2$.  The justification for neglecting scalar effects is the
same as before; for more discussion see 
footnote \ref{rant}.

\begin{table}[t]
\centerline{
\begin{tabular}{|c|c|c|c|}\hline
Potential used & Wave Function $Z$ & $\quad -\delta m^2/g^4 T^2 \quad$
& $\quad \sigma / g^4 T^3 \quad $ \\ \hline
1 Loop              & Tree            & 0.0058 & 0.0302 \\ \hline
                    & Tree            & 0.0168 & 0.088 \\ 
2 Loop              & $Z_{\rm exp}$   & 0.0109 & 0.072 \\ 
                    & $Z_{\rm pade}$  & 0.0115 & 0.074 \\ \hline
                    & Tree            & 0.0151 & 0.097 \\
``nonperturbative'' & $Z_{\rm exp}$   & 0.0110 & 0.083 \\
                    & $Z_{\rm pade}$  & 0.0114 & 0.084 \\ \hline \hline
Nonpert. Result     &                 & 0.0084 & 0.075 \\ \hline
\end{tabular}}
\vspace{0.2in}
\caption{\la{veff_table} Supercooling and surface tension, using
several analytical or semi-analytical approaches, compared with the full 
nonperturbative answer (last line).  The meanings of $Z_{\rm pade}$ and
$Z_{\rm exp}$ are explained in the next subsection.  The one loop
result gives too small results; all other results give too large an
answer.}
\end{table}

Including this term in the effective potential, we find that, while
$\phi_0^2 = 2.19$ is still smaller than the nonperturbative $\Delta
\avphi$, we now get too large a surface tension and too much
supercooling, by a substantial margin:  
\begin{equation}
\sigma_{2 \; {\rm loops}} = 0.088 g^4 T^3 \, , \qquad
\delta m^2({\rm 2 \; loops}) = -0.0168 g^4 T^2 \, .
\end{equation}
The amount of supercooling is too large by a factor of 2.
At the value of $\delta m^2$ found nonperturbatively, 
$S_{\rm 2 \; loop}>260$ is almost 3 times too large.
All the results for various effective potentials are summarized in Table 
\ref{veff_table}.

It is also possible to define a nonperturbative ``effective potential
for $\phi$,'' as follows.  We first find the largest volume for which
the configurations with $\avphi$ intermediate between the pure phase
values, do not show phase segregation.  This turns out to be a
surprisingly large volume.  To see this, recall why, in a larger box, a
configuration with $\avphi$ intermediate between homogeneous phase
values is a mixed phase configuration, 
rather than an extensive region with the
intermediate field value.  The effective potential we will eventually
derive is shown in Fig.~\ref{Veff_nonpert}.  The state half way between
minima has negative curvature.  Infrared Higgs field fluctuations are
therefore spinodally unstable.  If we force $\avphi$ to maintain its
value, this prevents the zero mode from growing, ie.~it keeps $\phi$ from 
shifting value homogeneously through the box; but small $k$ modes will
be unstable.  Any mode with $k < \omega_-$, with $\omega_- =
\sqrt{-V''}$ evaluated at the unstable point in question, will be
unstable to grow.  However, in a finite volume, there is a discrete
spectrum of $k$ modes.  The lowest nontrivial mode has $k = 2 \pi / L$;
so in any volume smaller than $L = 2 \pi / \omega_-$,
configurations with intermediate $\avphi$ will be homogeneous.  In
larger boxes they will be inhomogeneous, containing a region closer to
one phase and a region closer to the other.

\begin{figure}[t]
\centerline{\epsfxsize=3in\epsfbox{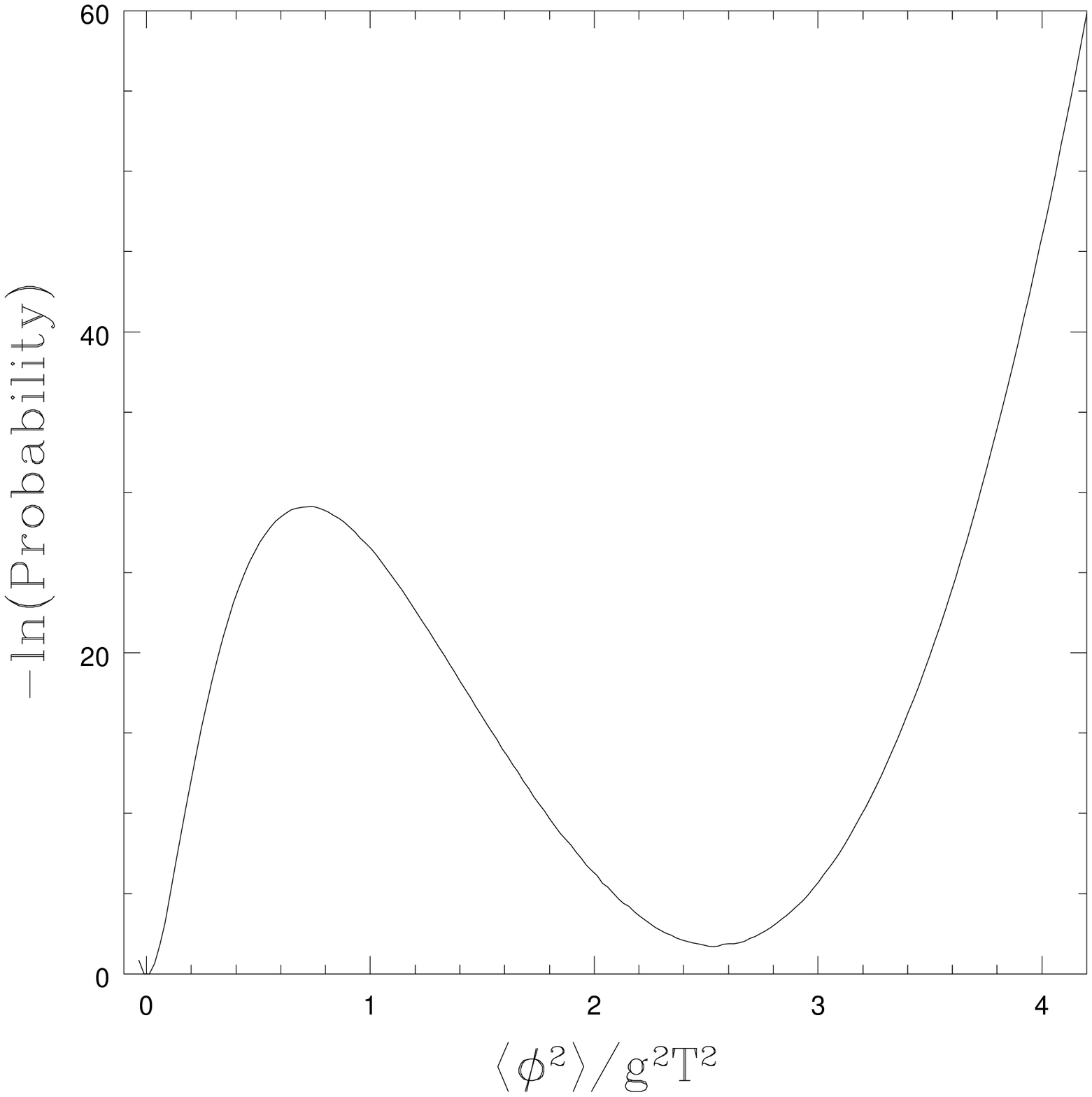} \hspace{0.4in}
\epsfxsize=3in\epsfbox{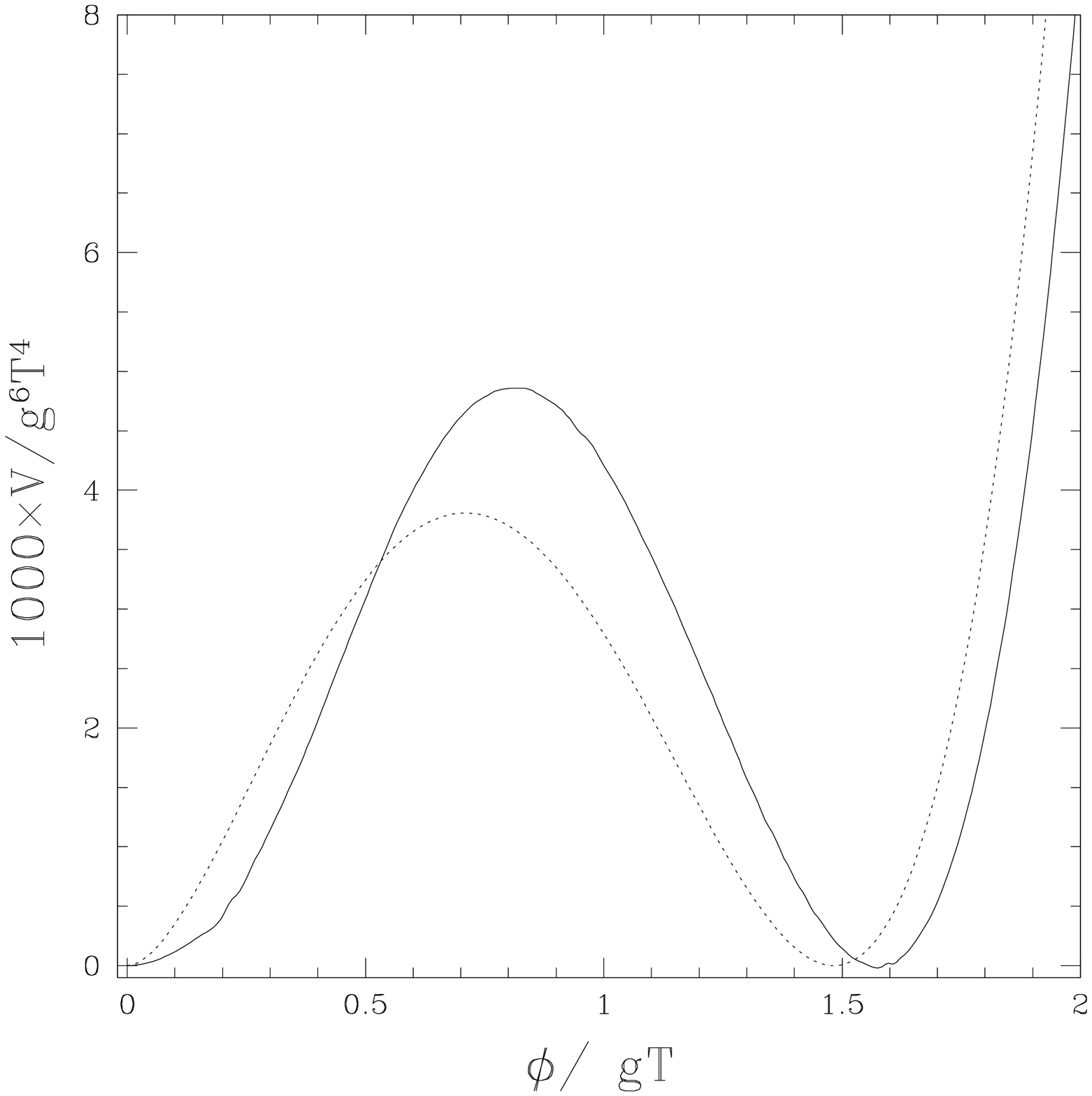}}
\vspace{0.2in}
\caption{\la{Veff_nonpert} Left:  free energy ($ - \ln({\rm
probability})$) distribution as a function of $\avphi$ in a $40^3$ box
at $a=4/9 g^2T$ ($\beta=9$).  The numerical errors are $<0.2$ and not
shown on the plot.  Right:  plot at left
interpreted as an effective potential for $\phi$, as described in the
text.  The dotted line is the 2 loop perturbative result, included for
comparison.}
\end{figure}

Now, we determine $F(\avphi)$, meaning $- \log({\rm Prob}(\avphi))$, in
such a ``large but not very large'' volume.  In practice we use a $40^3$ 
lattice at $\beta=9$, meaning a volume $17.8/g^2T$ on a side.  This is
pretty big, though the volumes we used to study critical bubbles were
typically 3 times longer on a side.
We then write $V = F/V$ and $\phi = \sqrt{ \avphi 
- \avphi({\rm s \; phase})}$, with $\avphi({\rm s \; phase})$ the lower
local minimum of $F(\avphi)$ at $\Teq$.  That is, we throw away the part 
of the picture on the left in Fig.~\ref{Veff_nonpert} which lies at
smaller $\avphi$ than the first minimum, and rescale the $x$ axis for
the rest.  We take the result to be a ``nonperturbative measured
$V(\phi)$'', shown at the right in the figure.

The advantages of getting an effective potential in this way are that it
should automatically get the right latent heat, and will show the
disappearance of the phase transition at large $\lambda / g^2$.  
The disadvantages are that it is numerically
expensive (though less so than a direct determination of the nucleation
rate), is somewhat arbitrary (how exactly do we choose a volume, for
instance?), and is still only approximative.  In particular, it is not
at all clear that it is reasonable to assume
that $\phi$ as defined above will have canonically normalized gradient
energies.

The ``nonperturbative effective potential'' has a larger separation
between minima, but the height of the barrier rises relative to the two
loop potential more weakly than as $\phi_0^4$.  As a result it gives a
slightly lower degree of supercooling, $\delta m^2 = -0.0151$, as
summarized in Table \ref{veff_table}.  Note however that the surface
tension is further off than in two loop perturbation theory.
What this exercise teaches us is that 
the difference between the perturbative and the nonperturbative
values of $\sigma$ and $\delta m^2$ are not primarily because of
the limitations of the 2 loop effective potential.

\subsection{Perturbative estimates including wave function corrections}

We have just found that estimates of the degree of
supercooling and of the surface tension were not substantially improved
by passing from the one loop to the two loop effective potential, using
the tree level Higgs field kinetic term in each case.  In fact this
result is not surprising.  While the two loop effective potential can
tell us the latent heat $l \propto \avphi$ at next to leading order in
$\lambda / g^2$, the procedure we used is still correct only at leading
order in $\lambda / g^2$ for determining the quantities we want, because 
it does not incorporate thermal corrections to the Higgs field wave
function.  In this subsection we see what happens when we incorporate Higgs 
wave function corrections which make the calculation complete to next to 
leading order in $\lambda / g^2$.

\begin{figure}
\centerline{\epsfxsize=4in\epsfbox{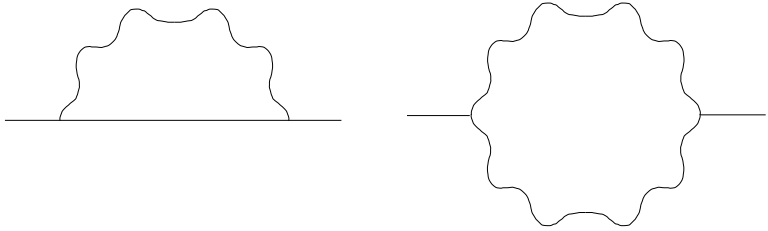}}
\vspace{0.2in}
\caption{\la{wave_fig} Diagrams which lead to $O(\lambda / g^2)$
important Higgs wave function corrections.  The vertices in the diagram
at right only exist because there is a condensate.}
\end{figure}

Using the one loop effective potential, we can get the parametric
estimate $\phi_0 \sim g^3 T/\lambda$, so $m_W(\phi) \sim g^4 T /
\lambda$ in the broken phase and inside the wall.  The wall thickness,
on the other hand, is set by the Higgs boson mass in either phase,
$L_{\rm wall} \sim 1/m_{HT} \sim g^3 T / \sqrt{\lambda}$.  Therefore, up 
to a correction suppressed by a half power of $\lambda / g^2$, we may
take the $W$ bosons to be heavy compared with the reciprocal wall width, 
and treat the Higgs condensate as a homogeneous background for them.
For this reason it is possible to incorporate the leading non-potential
correction as a wave function correction for the Higgs field, 
\begin{equation}
\frac{1}{2} (\nabla \phi)^2 \quad \Rightarrow \quad
	\frac{Z(\phi)}{2} (\nabla \phi)^2 \, .
\end{equation}
Expanding the diagrams shown in Fig.~\ref{wave_fig} to second order in
external momenta (treated as much less than $m_W$), we find a correction 
to the Higgs gradient term which is, in Landau
gauge,
\begin{equation}
Z(\phi) = 1 + \delta Z(\phi) = 1 - \frac{11}{32 \pi} \frac{g^2 T}{m_W} 
	= 1 - \frac{11}{16 \pi} \frac{gT}{\phi} \, ,
\end{equation}
in agreement with the expression found by B\"{o}deker
et.~al.~\cite{Bod_et_al} (see also
\cite{LaserSchmidt}, where these wave function corrections have been
considered at length).

Both the corrections from Higgs fields, and from
higher order in the $p \ll m_W$ expansion, will give 
at most $O(\lambda^{3/2}/ g^3)$ corrections to $\sigma$ and $\delta
m_{HT}^2$; whereas we see from the estimate for $\phi_0$ that the above
term contributes at $O(\lambda / g^2)$.  It will drive down both
$\sigma$ and $\delta m^2$, since it lowers the gradient
energy contribution to the energy, and so makes it cheaper to have
bubbles or interfaces.  We should mention that to compute the $(\lambda
/g^2)^{3/2}$ corrections, we would have to perform a fluctuation
determinant, since the Higgs field mass and the width of the interface
are parametrically the same.\footnote{Such a fluctuation determinant
calculation has been performed by Baacke
\pcite{Baacke}, and has also recently been considered by
Parnachev and Yaffe \pcite{Parnachev}; 
but since neither reference uses the two loop
effective potential, their results cannot be considered correct through
to $O(\lambda /g^2)$.  It is not clear to us how  
to simultaneously perform the fluctuation determinant and to incorporate 
the two loop gauge contributions to the effective potential.  Without
resolving this question, the calculation of the fluctuation determinant
is not justified in the sense of an expansion in $\lambda / g^2$.}

Unfortunately $Z(\phi)$ goes to zero at a finite value of $\phi$.  At
the value of $\phi$ where this happens, the condensate is so weak that
perturbation theory is also breaking down; the failure signals that
neglected higher order effects become essential.  To deal with this, we
will make an {\it Ansatz} for those effects, chosen to maintain the
correct large $\phi$ behavior of $Z(\phi)$ but to prevent $Z(\phi)$ from 
going to zero.  We have considered two
choices; to approximate
\begin{equation}
Z_{\rm exp} = \exp ( - \delta Z ) \, , \qquad {\rm and} \qquad
Z_{\rm pade} = \frac{1}{1 - \delta Z} \, .
\end{equation}
Since we do not know the higher order behavior of $Z$, we do not know
which of these is more sensible.  If the answers we get depend strongly
on which one we take, that is an indication that the perturbative
expansion is failing and we cannot trust either.

When we re-compute $\delta m^2$ and $\sigma$, including these wave
function corrections, we get a result which is much closer to the
nonperturbative value, see Table \ref{veff_table}.  Further, the choice
of how to resum higher terms in $Z(\phi)$ appears not to matter much.
However, the supercooling is still over-estimated by about $25\%$.
At $\delta m^2$ where the nonperturbative calculation shows the phase
transition takes place, where $S_{\rm nonpert}=90$,
the 2 loop potential with Pad\'e resummed self-energy corrections
gives $S=143$, more than $50\%$ high.

In summary, using a perturbatively computed effective potential but tree 
Higgs kinetic terms appears to
work badly.  Higgs wave function corrections are
important and improve the performance of the perturbative treatment;
they should be included in any subsequent work which tries to
study electroweak bubble nucleation by perturbative means.  However,
even with wave function corrections, perturbation theory is still not a
very accurate way to treat bubble nucleation.

%------------------------------------------------------------------
\section{Conclusions}
\la{Conclusion}

In this paper we have presented a technique for determining bubble
nucleation rates in theories with classical infrared thermodynamics and
dynamics, in a fully nonperturbative way--even where the rate is
exponentially small.  The technique can be considered a generalization
of Langer's formalism \cite{Langer}, which replaces the approximate
saddle point expansion of that method with a nonperturbative
Monte Carlo calculation, and takes care to treat correctly the
microscopic dynamics of the nucleation process.  The procedure uses an
interesting mixture of multicanonical Monte Carlo and real time
techniques.  Within the context of the dimensional reduction
approximation for the thermodynamics, and B\"{o}deker's effective theory 
for the dynamics \cite{Bodeker}, our treatment is exact up to small and
controllable numerical errors.  It is useful 
when the microscopic physics is known well enough, and suitably amenable 
to a lattice treatment, to permit an accurate first principles
Monte Carlo calculation.  In particular, it can be applied at the
electroweak phase transition.

We have applied the technique to the electroweak phase
transition in the standard model, at a value of the coupling which is
just enough to preserve any baryon number after the transition; the
ratio of (3-D effective theory couplings) was taken as $\lambda
/g^2=0.036$.  The degree of supercooling is such that the thermal Higgs
mass squared falls by $\delta m^2 = -0.00840 g^4 T^2$ from its
equilibrium value.  

The main value of 
this measurement is that we can compare it to the results of more
traditional and less first principles calculations.  We find that the
most common method in the recent literature, using the two loop
effective potential but tree level Higgs kinetic term, is very
unreliable.  For the parameters considered it overestimates the amount
of supercooling by a factor of 2, even though it gets $\phi_0$, the
length of the Higgs field in the broken phase, to within
$10\%$ error.  If one considers the {\em action} of the critical bubble, 
it is even {\em further} off.  
Including Higgs field wave function corrections
substantially improves the accuracy; the amount of supercooling is then
only overestimated by about $25\%$.  The remaining discrepancy cannot be 
attributed to the inaccuracy of the effective potential; a
nonperturbative ``effective potential for $\phi$'' is off by the same
amount.  On the other hand, a thin wall estimate, with nonperturbative
inputs, does quite well--it is off by $20\%$, in the opposite direction.

It should be straightforward to apply our technique to the more
phenomenologically interesting MSSM or NMSSM, which can support viable
baryogenesis. 

\centerline{\bf Acknowledgments}

We thank Mikko Laine for many useful discussions.
KR acknowledges partial support from EU TMR grant FMRX-CT97-0122.
A part of the simulations have been run on the Cray T3E at the Center for
Scientific Computing, Espoo, Finland.

\appendix

\section{Discussion:  independence on choice of measurable}
\la{separatrix}

In this appendix we justify why the procedure presented in Section
\ref{Plan} of the main text
``works,'' and in particular why the determined rate should be
independent of the choice of measurable.

Fix attention to a particular volume, and a particular 
value of $T < T_{\rm eq}$, at which the
nucleation rate is small.  We must begin by clarifying the {\em
definition} of the nucleation rate.  First, we must be able to
say whether or not a configuration is in the metastable phase.  This
requires that we possess {\em some} measurable in terms of which the free
energy shows a ``two well'' structure, like Fig.~\ref{maxwell3}, with
the probability at $(C)$ exponentially smaller than at $(A)$.  For
simplicity of notation let us assume that $\avphi$ serves this purpose.

We can now draw two lines, roughly at $(B)$ and $(D)$ in
Fig.~\ref{maxwell3}; all we require is that they be well between $(C)$
and the local minima, such that the likelihood to be at $(B)$ is
exponentially greater than the likelihood to be at $(C)$, but
exponentially less likely than to be at $(A)$ (and similarly $(D)$ is
exponentially more likely than $(C)$ but less likely than the broken
phase).  We
say that a configuration is ``definitely in the metastable phase'' if it
has a value of $\avphi$ to the left of $(B)$, and is ``definitely in the
broken phase'' if it has a value to the right of $(D)$.

The intuitive meaning of the nucleation rate is the following.  Take the
canonical ensemble, and throw out all the configurations to the right of
$(B)$, leaving only the ones which are clearly in the metastable
minimum.  Now carry out the time evolution for a ``medium'' amount of
time, $t_{\rm medium}$, exponentially
longer than any microscopic time scale, but exponentially shorter than
the time it takes for most of the metastable configurations to escape to
the broken phase.  At the end of the time evolution, look to see how
much of the ensemble is definitely in the broken phase, ie.~to the right
of $(D)$.  Define that fraction, divided by $V t_{\rm medium}$, to be
the spacetime rate of nucleations.

This intuitive meaning of the nucleation rate will be our definition.
Note that it is equivalent to the following.
Consider the set of all trajectories of length $t_{\rm medium}$,
starting from any configuration (symmetric, broken, or in between) with
appropriate Boltzmann weight.  The nucleation rate per unit volume is
\begin{equation}
{\rm rate} = \frac{ {\rm P}({\rm symm} \rightarrow {\rm broken})}
	{V t_{\rm medium} {\rm P}({\rm symm} \rightarrow 
	{\rm symm})} \, ,
\end{equation}
where (${\rm P}$) means the fraction of the trajectories satisfying
the given condition.
In other words, find what fraction of trajectories start in the
symmetric phase and end in the broken one; and divide by the number
which start and end in the symmetric phase, the volume, and the time.
(The denominator should read ${\rm P}({\rm symm} \rightarrow {\rm
left \; of \;}(D))$, but since configurations between $(B)$ and $(D)$
are exponentially rare the difference is exponentially small.)
This definition of a rate depends on our exact choices for $(B)$, $(D)$,
and $t_{\rm medium}$, but the sensitivity should be exponentially weak
--- unless there is some additional long time scale in the problem,
besides the nucleation rate, in which case our technique probably fails.
We can think of the
nucleation problem in terms of sampling over the space of real time
trajectories, and the sampling may be taken using the equilibrium
ensemble.  
The goal is now to show that our technique correctly determines the 
number of crossing trajectories, relative to all trajectories which
remain near the symmetric phase; and that this does not depend
sensitively on our choice of measurable.

{}From here on we will consider our technique applied to two classes of
dynamics.  The first is Hamiltonian dynamics, for a system where the
phase space is the tangent bundle
over the space of configurations.  We further assume that the
Hamiltonian is a function of space, plus (a constant times) the inner
product function on the tangent space.  In other words, the momenta
appear quadratically in the Hamiltonian, and in any orthonormal basis for
the momenta the quadratic form is a multiple of the identity.  Probably
these conditions are too strong, and we could work with any Hamiltonian
system which could be written as a fiber bundle over a configuration
space; but the treatment would be more complicated.  The other class of
dynamics we consider is Langevin dynamics, where the noise term is
Gaussian, white, and additive.  White means there are no unequal time
correlators in the noise, in which case it is described by a weight
function on the tangent space; we take that weight function to be of the
same form as the momentum distribution just described for the
Hamiltonian case.
Probably one could extend our technique to the case
where the noise is stronger in some coordinate directions, or has an
amplitude which varies in space.  However, such Langevin systems are
notoriously subtle, see eg.~\cite{ASY_subtle,Arnold_Langevin}.

Consider two measurables, call them ${\cal O}_1$ and ${\cal O}_2$.
(They could for instance be $\avphi$ and some other volume average of a
local observable; or ${\cal O}_2$ could for instance be the maximum over
all centerpoints for spheres of radius $r$, of $\int_{\rm sphere}
2\Phi^\dagger \Phi$.)  For each observable we can make a free energy
plot and identify some least likely value of ${\cal O}$, ${\cal O}_{\rm
crit}$.  The measurable ${\cal O}$ is
a map from the space of configurations to the real
numbers; the subspace which gives a particular value of ${\cal O}$ is a
codimension 1 surface in the space of configurations.  Hence defining a
measurable also defines a foliation of the configuration space.  We will
call the special surface, consisting of all configurations with ${\cal
O} = {\cal O}_{\rm crit}$, the {\em separatrix} for the measurable ${\cal
O}$.  This notation is in keeping with \cite{KhlebShap}, where many of
the root ideas of our algorithm can be found.
One would like the separatrix to separate the configurations more
likely, under time evolution, to go to the symmetric phase, from those
more likely to go to the broken phase.  

\begin{figure}[t]
\centerline{\epsfxsize=3.5in\epsfbox{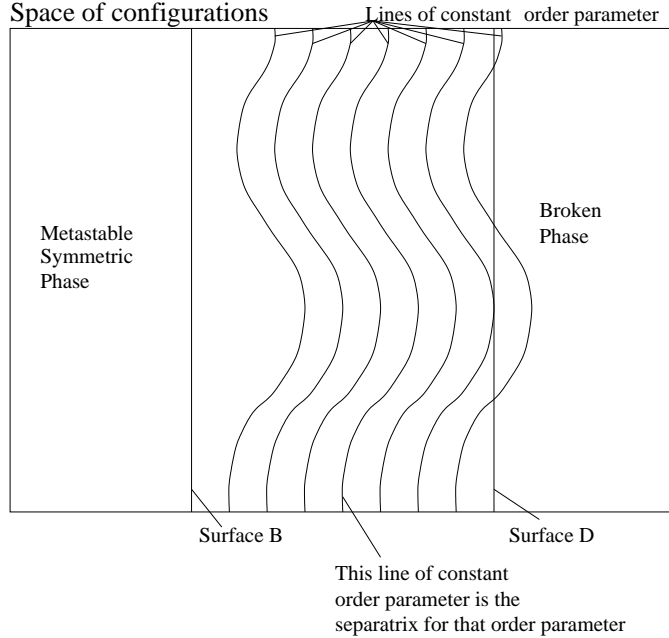}}
\caption{A cartoon of how configuration space is foliated by a
measurable--here the lines of constant measurable do not coincide with
the ones which determine the surfaces $B$ and $D$.\la{sep1}}
\end{figure}

Note that $(B)$ and $(D)$ also give codimension 1 surfaces
in the configuration space, namely the surfaces where $\avphi$ takes on
the values $(B)$ and $(D)$; we will call these surfaces $B$ and $D$.  
To one side of the surface $B$ is the
metastable phase, to the far side of surface $D$ is the broken phase,
and in between are intermediate configurations.  For the measurable
${\cal O}$ to be useful, we require that the ${\cal O}$ separatrix carry
all but an exponentially small part of its weight between the surfaces
$B$ and $D$.  We give a cartoon of how these surfaces in configuration
space look, in Fig.~\ref{sep1}.

It is possible to define an ideal operator, ${\cal
O}_{\rm ideal}$, which will provide an ideal separatrix for
distinguishing configurations which are in the domain of attraction of
one or the other phase.  Namely, we define
\begin{equation}
{\cal O}_{\rm ideal}({\rm config}) = \int_{\rm traj. \; through \;
config} \Theta(\avphi({\rm config}(t_{\rm medium}))<(B)) \, .
\end{equation}
${\cal O}_{\rm ideal}$ of a configuration is the
fraction of trajectories, starting at that configuration, which are in
the metastable phase after time $t_{\rm medium}$.  For
Hamiltonian dynamics, the measure on the space of trajectories through a
configuration is just the canonical measure of the tangent space; a
point in the tangent space uniquely defines a trajectory.  For Langevin
dynamics, the measure for the space of trajectories through a point is
given by the measure of realizations of the noise, with each trajectory
corresponding to the noise realization which generates it.
We do not use ${\cal O}_{\rm ideal}$ in our work because its measurement
is impractical.

The method presented in the main text for the determination of the
nucleation rate, applied to an observable ${\cal O}$, can be phrased as
follows:  first, we find the probability distribution as a function of
the value of ${\cal O}$; that is, we find the integral along each
surface of constant ${\cal O}$ of the weight of the canonical
ensemble.  This gives
\begin{equation}
 \frac{	{\rm P}(|{\cal O} - {\cal O}_{\rm crit} | < \epsilon/2)}
	{ \epsilon \: {\rm P}({\cal O} < {\cal O}_{\rm crit})}
	\propto
	\int_{\rm surf} d({\rm surf.\; area}) \exp(-H/T) 
	\left[ d {\cal O}/d({\rm normal}) \right]^{-1} \, ,
\end{equation}
the integral over the surface of the Boltzmann weight, {\em divided by}
the surface normal derivative of ${\cal O}$.  The surface normal
derivative appears because, the
larger the derivative is, the narrower is the region over which
${\cal O}$ differs by less than $\epsilon/2$ from ${\cal O}_{\rm
crit}$.  

Multiplying by $\langle |\Delta {\cal O}/\Delta t| \rangle$ precisely
compensates for this normal derivative factor.  This is because the mean
value of $|\Delta {\cal O}/\Delta t|$, at any point in configuration
space, is proportional to the gradient of ${\cal O}$ at that
point.  To see this, note first that only motion normal to the surface
of constant ${\cal O}$ matters.  Next note that, by our requirements on
the dynamics, the
rms.~metric distance traveled in one direction, on averaging over
possible momenta (or Langevin noise realizations), 
is independent of position in configuration space or direction
considered.  Therefore
\begin{equation}
\left\langle \left| \frac{ \Delta {\cal O}}{\Delta t} \right| 
	\right\rangle \propto \frac{ 
	\int_{\rm surf} d({\rm surf.\; area}) \exp(-H/T)}
	{\int_{\rm surf} d({\rm surf.\; area}) \exp(-H/T)
	\left[ d {\cal O}/d({\rm normal}) \right]^{-1}} \, .
\end{equation}
The product is proportional to the Boltzmann weighted area of the
separatrix, and hence the flux through the separatrix, as claimed in the
body of the paper.

\begin{figure}
\centerline{\epsfxsize=4in\epsfbox{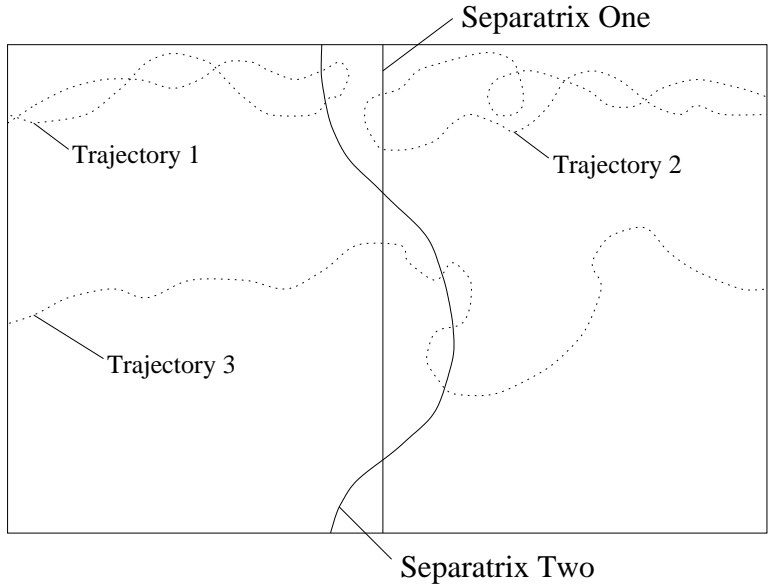}}
\vspace{0.1in}
\caption{cartoon of separatrices for two measurables, and some
trajectories.  Trajectory 1 crosses separatrix 2 but not separatrix 1;
but it (therefore) crosses an even number of times, and leads to a 0
entry in determining $\d_{{\cal O}2}$.  Similarly Trajectory 2
crosses separatrix 1 but not separatrix 2, but also does not contribute
to $\d_{{\cal O}1}$.  Trajectory 3 does lead to a nucleation.  It is
sampled 3 times in computing $\d_{{\cal O}2}$, each time
contributing (1/3); and sampled once in computing $\d_{{\cal O}1}$,
contributing 1.  Hence it gives the same contribution to the total rate
computation for each measurable.
\la{sep2}}
\end{figure}

The flux through the separatrix will clearly differ for different
choices of ${\cal O}$, as illustrated in Fig.~\ref{sep2}.
It remains to show that $\d_{\cal O}$, as defined in
Eq.~\ref{eq_for_d}, precisely turns the correctly
normalized flux of trajectories through the separatrix into a count of
trajectories which mediate nucleations.  Fig.~\ref{sep2} illustrates
why this is true.  First consider a trajectory which crosses the
separatrix an even number of times and returns to the phase it came
from, like Trajectory 1 in the figure.  If the trajectory crosses $n$
times, it gets counted $n$ times in the sampling procedure, once at each
crossing of the separatrix.  Each count is with the same weight.  For
Hamiltonian dynamics this is because the evolution conserves energy
(hence the Boltzmann factors are all the same) and phase space measure.
For Langevin dynamics it is because the Langevin dynamics correctly
generate the thermal ensemble.  In either case, the number of times the
trajectory gets sampled is in proportion to its contribution to the
flux, and each time it contributes zero to $\d$; hence $\d$
correctly accounts for the number of nucleations (zero) the trajectory
causes.  

Next consider a trajectory which does get from the metastable phase to
the stable one.  It is guaranteed to cross both separatrices an odd
number of times, because each is to the right of $B$ where the
trajectory starts, and to the left of $D$ where it ends.  If the
trajectory crosses a separatrix $n$ times, it will appear in the
average, used to determine $\d$, $n$ times, each with the same
weight; and each appearance contributes $1/n$ to the determination of
$\d$, so the number of nucleations is correctly counted as 1.
Again, the Boltzmann weighted amount of flux the trajectory crossing
represents, is the same at each of its crossings --- of either
separatrix.  Hence if there is a larger total flux through, say, the  
separatrix for ${\cal O}_1$, then this individual trajectory represents
a smaller fraction of that flux, and gets an appropriately smaller
weight in the sampling for determining $\d_{{\cal O}1}$.  Its
positive contribution to the value of $\d_{{\cal O}1}$ is
correspondingly smaller.
Since the set of trajectories which mediate nucleations are the same,
whichever separatrix we use to sample them, 
the smaller size of $\d$ exactly compensates for the larger flux
through the separatrix.  This is why the total rate computed is the same
for each choice of ${\cal O}$.

While different observables give
the same answer for the nucleation rate, they are not equally convenient
as numerical tools.  A good measurable must satisfy two conditions.
First, it must be easy to measure it, so it can be used practically in
reweighting configurations in the Monte Carlo (see subsection
\ref{multi_subsec}).  Second, when determining $\d$, statistics
must accumulate with a reasonable sample of configurations, which means
that a reasonable fraction of trajectories through the separatrix must
actually lead to bubble nucleation.  Otherwise it may take an
exponentially large sample of trajectories to determine ${\bf
d}$ with good statistical accuracy.  Roughly, this will require that the
separatrix of the observable is close to degenerate with the separatrix of
${\cal O}_{\rm ideal}$.  Note that, by construction, under Langevin
dynamics half of trajectories through the ${\cal O}_{\rm ideal}$
separatrix lead to nucleations, and that this is the upper bound among
all observables.

\newcommand{\hep}[1]{[{#1}]}

\end{document}